\documentclass[fleqn,usenatbib]{mnras}

\usepackage[T1]{fontenc}
\usepackage{ae,aecompl}

\usepackage{graphicx}	% Including figure files
\usepackage{amsmath}	% Advanced maths commands
\usepackage{amssymb}	% Extra maths symbols
\usepackage{bm}
\usepackage{comment}
\usepackage{newtxtext,newtxmath}

\title[Simulations of wind-reprocessed transients]{Moving-mesh radiation-hydrodynamic simulations of wind-reprocessed transients}

\author[D. Calder\'on et al.]{
Diego Calder\'on$^{1}$\thanks{E-mail: diego.calderon@utf.mff.cuni.cz (DC)},
Ond\v{r}ej Pejcha$^{1}$,
Paul C. Duffell$^2$
\\
$^{1}$Institute of Theoretical Physics, Faculty of Mathematics and Physics, Charles University, V Hole\v{s}ovi\v{c}k\'ach 2, 180 00, Praha 8, Czech Republic\\
$^{2}$Department of Physics and Astronomy, Purdue University, 525 Northwestern Avenue, West Lafayette, IN 47907-2036, USA
}

\date{Accepted XXX. Received YYY; in original form ZZZ}

\pubyear{2021}

\begin{document}
\label{firstpage}
\pagerange{\pageref{firstpage}--\pageref{lastpage}}
\maketitle

\begin{abstract}
    Motivated by recent theoretical work on tidal disruption events and other peculiar transients,
    we present moving-mesh radiation-hydrodynamic simulations of radiative luminosity emitted by a central source being reprocessed by a wind-like outflow. 
    We couple the moving-mesh hydrodynamic code JET  with our newly-developed radiation module based on mixed-frame grey flux-limited diffusion with implicit timestep update. 
    This allows us to study the self-consistent multi-dimensional radiation-hydrodynamic evolution over more than ten orders of magnitude in both space and time in a single run. 
    We simulate an optically-thick spherical wind with constant or evolving mass-loss rate, which is irradiated by a central isotropic or angularly-dependent radiation source. 
    Our spherically-symmetric simulations confirm previous analytic results by identifying different stages of radiation reprocessing: radiation trapped in the wind, diffusing out through the wind, and reaching constant maximum attenuation. 
    We find that confining the central radiation source in a cone with moderate opening angles decrease up to one order of magnitude the early flux along sightlines oriented away from the direction of radiation injection but that the reprocessed radiation becomes isotropic roughly after one lateral diffusion time through the ejecta. 
    We discuss further applications and guidelines for the use of our novel radiation-hydrodynamics tool in the context of transient modelling.
\end{abstract}

\begin{keywords}
transients -- radiative transfer -- methods: numerical -- radiation: dynamics
\end{keywords}

%%%%%%%%%%%%%%%%%%%%%%%%%%%%%%%%%%%%%%%%%%%%%%%%%%

%%%%%%%%%%%%%%%%% BODY OF PAPER %%%%%%%%%%%%%%%%%%

\section{Introduction}

    There is mounting evidence that new sources of internal power or non-spherical geometry are required to explain certain astronomical transients.
    For example, a fraction of the observed core-collapse supernovae can only be explained as a spherical explosion colliding with an non-spherical distribution of circumstellar medium (CSM) \citep[e.g.][]{chugai94,leonard00,smith15,bilinski18}. 
	In some of these situations, the ensuing radiative shocks can be engulfed by the nearly spherical ejecta and act as an internal power source \citep[e.g.][]{mauerhan13,smith13,andrews18}. 
	These observations have motivated theoretical investigations of supernova explosions colliding with oblate and prolate CSM distributions such as disks, bow shocks, colliding-wind shocks, or bipolar nebulae \citep{blondin96,vanmarle10, vlasis16,S16,S19,mcdowell18,kurfurst19,kurfurst20}. 
	
	In the case of classical novae, high-resolution radio imaging of an expanding shell in V959~Mon has shown distinct equatorial and polar components separated by a shock interaction region \citep{chomiuk14}. 
	Furthermore, the recently found correlations between optical and GeV $\gamma$-ray variability suggest that non-spherical shock interaction might be actually powering the optical light curves \citep[e.g.][]{metzger14,li17,aydi20}. 
	
	The discovery of new transient classes has also led to the development of detailed theoretical models typically involving the presence radiative shocks, interaction with CSM, deviations from spherical symmetry, and/or radiation-reprocessing outflows. 
	For instance, stellar mergers and common envelope events have been associated with luminous red novae and intermediate luminosity optical transients \citep{soker03,tylenda06,ivanova13}. 
	These events could be powered by a radiative shock embedded in a hydrogen-rich ejecta, where the different ejecta and CSM components reflect the runaway increase of mass loss from the binary star approaching the dynamical phase of the binary-star interaction \citep{pejcha14,pejcha17,metzger17,blagorodnova21}. 
	
    Among other enigmatic transients, observations of AT2018cow from the recently identified class of fast-rising blue optical transients revealed an X-ray source, likely a compact object or an embedded radiative shock, reprocessed by a non-spherical CSM \citep{margutti19,fang20,uno20}. 
    On a similar note, tidal disruption events inherently require multi-dimensional modelling as different signatures might be observed depending on the line-of-sight orientation. 
    Some of the salient features of tidal disruption events could be jets as well as radiation reprocessing in an outflow with polar-angle dependent velocity \citep[see][for a review]{D21}. 
    The radiation reprocessing by an outflow is necessary to explain the observed peak in optical or UV frequencies rather than the expected EUV or soft X-ray. 
    This fact led to the development of models for the general problem of how radiation is reprocessed in optically-thick winds \citep[e.g.][]{shen16,uno20,P20}. 
    Such models have been used to explain the observed light curves of tidal disruption events by constraining the mass-loss rate of the winds as well as the spatial scale from where it is being launched \citep[e.g.][]{metzger16,uno20b}. 
    Interestingly, radiation reprocessing by optically-thick winds is also relevant for classical novae \citep[see][for a review]{chomiuk2020}. 
    So far, these theoretical models have not been verified with numerical time-dependent simulations, which we aim to address in this paper.
    
    Numerical radiation hydrodynamics is challenging and computationally expensive, especially in multi-dimensional models. 
	Among the numerical tools specialised for transient phenomena, one-dimensional Lagrangian approach is popular for quick simulations of light diffusion in expanding matter and for computing light curves. 
	A few examples of such codes are STELLA, which follows a multigroup radiative transfer approach \citep{blinnikov98}, SNEC \citep{morozova15}, which treats radiation under the grey flux-limited diffusion (hereafter FLD) approach \citep{A73}, and others of similar nature \citep{bersten11,pumo11}. 
	There are also codes for computing spectra from (radiation-)hydrodynamic simulations, e.g. SEDONA, which follows a Monte Carlo approach for solving the multi-frequency  three-dimensional time-dependent radiative transfer problem \citep{kasen06}, CMFGEN, which can synthesise light-curves and spectra for systems not necessarily in local-thermal equilibrium \citep[e.g.][]{dessart05}, TARDIS, which performs fast calculations of spectra for one-dimensional transient models \citep[e.g.][]{kerzendorf14}, and others. 
	These spectral codes differ by the level of coupling radiation and hydrodynamics ranging from operating on a pre-determined background to full coupling to one- or multi-dimensional hydrodynamic codes \citep[e.g.][]{roth15}.
	
	There are also multi-dimensional finite-volume hydrodynamic codes with modules for treating and coupling radiation with hydrodynamics, for example RAGE, which uses a radiation diffusion algorithm under the grey approximation \citep{gittings08}, CASTRO, which includes a module for multi-group FLD \citep{Z11,zhang13}, HERACLES, which treats radiation through M1 closure \citep{gonzalez07,dessart19}, FLASH, which makes use of grey FLD for including radiation \citep{C19}, and more recently also grey FLD in AMRVAC \citep{moens21}. 
	Although these codes can achieve high spatial resolution thanks to their adaptive-mesh refinement modules, the finite-volume approach limits their domain size. 
	Despite these achievements, to our knowledge, \emph{there is no tool that combines the wide temporal and spatial dynamical range of one-dimensional Lagrangian codes with the ability to study transients lacking spherical symmetry}.
	This lack of capability limits comparison of theoretical models with observational data in a quick and straightforward manner.

    In this work, we perform two-dimensional moving-mesh radiation-hydrodynamic simulations of the radiation of a central source being reprocessed by an optically-thick spherical wind. 
    To do so, we have developed a new radiation module based on the mixed-frame formulation under the grey FLD approximation for the moving-mesh hydrodynamic code JET \citep{D13}. 
	JET was derived from the code TESS \citep{D11}, which uses a numerical mesh built through Voronoi tessellation of the computational domain. 
	However, as JET is intended to be used for modelling problems with rapid radial outflows, it instead uses spherical coordinates to construct a grid that is only allowed to move radially. 
	Thus, the code effectively acts as a set of one-dimensional Lagrangian codes that are coupled laterally by transverse fluxes. 
    The moving-mesh nature of the code helps to resolve contact discontinuities along the radial direction to high precision and, at the same time, aids to loose the constraints on the timestep. 
    Such properties make JET with our new radiation module an ideal tool for simulating astronomical transients, even for problems lacking of spherical symmetry.

	This work is organised as follows. 
	In Section~\ref{sec:analytical} we briefly summarise the analytical formalism for characterising wind-reprocessed transients. 
	In Section~\ref{sec:numerical}, we describe our numerical method including the approach used for the radiation treatment. 
	A more thorough description of the algorithm, method of solution, and consistency tests is presented in Appendices~\ref{sec:app1},~\ref{sec:app2}, and~\ref{sec:app3}, respectively. 
	The results of our simulations are presented and described in Section~\ref{sec:results}. 
	Finally, we present the conclusion of this work and future outlook in Section~\ref{sec:conclusions}.

\section{Analytical formalism}
\label{sec:analytical}
	
\begin{figure}
	    \centering
	    \fbox{\includegraphics[width=0.45\textwidth]{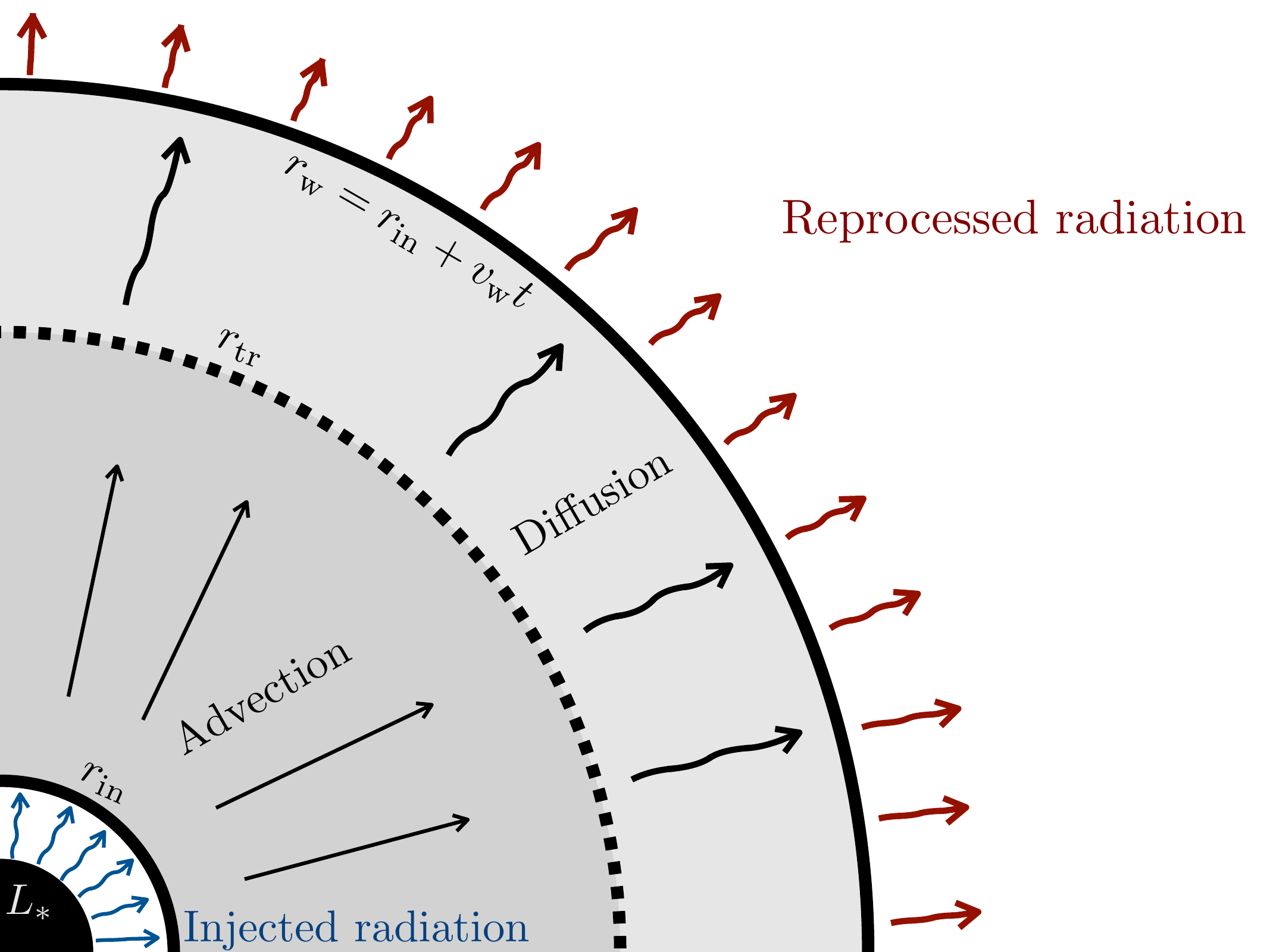}} 
	    \caption{
	    Schematic representation of a wind-reprocessed transient. 
	    The problem considers a central source launching a spherically symmetric wind at a radius $r_{\rm in}$ with a constant speed $v_{\rm w}$. 
	    After an arbitrary time $t$ the outer radius of the wind is $r_{\rm w}=r_{\rm in}+v_{\rm w}t$.
	    Simultaneously, there is a central source of luminosity $L_*$ that irradiates into the inner boundary $r_{\rm in}$.
	    The wind is assumed to be dense enough so that radiation is mainly advected up to a radius $r_{\rm tr}$. 
	    At larger radii, radiation can diffuse outwards.
	    } 
	    \label{fig:diagram1}
	\end{figure}

	\cite{P20} developed an analytical approach for estimating the amount of radiation reprocessed by an optically-thick spherically symmetric wind as a function of time. 
	Figure~\ref{fig:diagram1} presents a schematic representation of the problem. 
	In this section, we give a brief summary of their calculations and assumptions. Our goal for the rest of the paper is to verify the theory of \citet{P20}.
	
	Let us consider a wind with constant mass-loss rate $\dot{M}$ and constant speed $v_{\rm w}$ that is ejected at an inner radius $r_{\rm in}$. 
	Then, its outer radius extension as a function of time is
	\begin{equation}
		r_{\rm w}(t) = r_{\rm in} + v_{\rm w}t.
	\end{equation}
	The density of such a wind is assumed to be 
	\begin{equation}
		\rho(r)=\frac{1}{4\pi}\frac{\dot{M}}{v_{\rm w}r^2}.
	\end{equation}
	If we consider constant opacity dominated by electron scattering, $k_{\rm s}=0.34\rm\ cm^2\ g^{-1}$, then the optical depth between an arbitrary radius and the outer radius of the wind can be estimated by
	\begin{equation}
		\tau(r) = \int^{r_{\rm w}(t)}_\text{r} k_{\rm s}\rho dr= \frac{k_{\rm s}\dot{M}}{4\pi v_{\rm w}}\left(\frac{1}{r}-\frac{1}{r_{\rm w}(t)}\right).
	\end{equation}
	Here, the theory assumes that the wind is dense enough so that radiation is initially trapped in it. 
	As a result, radiation is mainly advected with the wind instead of diffusing (see Figure~\ref{fig:diagram1}). 
	In order to quantify the degree of radiation trapping, \cite{P20} introduce the dimensionless parameter $A$:
	\begin{equation}
		A=\frac{k_{\rm s}\dot{M}}{4\pi r_{\rm in}c},
		\label{eq:A}
	\end{equation}
	where is $c$ the speed of light. 
	It is important to remark that radiation is trapped in the wind only if the wind properties, mass-loss rate, and opacity are such that $A\gg1$. 
	
	We can estimate the characteristic timescales involved in the problem. 
	The photon diffusion timescale $t_{\rm dif}$ is defined as the time it takes for light at a given radius to travel up to the outer extension of the wind,
	\begin{equation}
		t_{\rm dif}=\frac{\tau(r)}{c}\frac{(r_{\rm w}(t)-r)r}{r_{\rm w}(t)}.
		\label{eq:tdif}
	\end{equation}
	The dynamical timescale $t_{\rm dyn}$ corresponds to the time necessary for a layer of wind at the inner radius to travel up to an arbitrary distance $r$,
	\begin{equation}
		t_{\rm dyn} = \frac{r-r_{\rm in}}{v_{\rm w}}.
		\label{eq:tdyn}
	\end{equation}
	Once these expressions are equal, $t_{\rm dif}=t_{\rm dyn}$, we can estimate the so-called photon trapping radius $r_{\rm tr}$. 
	Beyond this radius the radiation is not trapped anymore and can easily diffuse outwards (see Fig.~\ref{fig:diagram1}).
	
	Based on these analytical considerations, the ratio of the observed luminosity $L_\text{obs}$ to injected luminosity $L_*$ is
	\begin{equation}
		\frac{L_{\rm obs}}{L_*}=\left(\frac{r_{\rm tr}}{r_{\rm in}}\right)^{-2/3}\left(1-\frac{1}{v_{\rm w}}\frac{dr_{\rm tr}}{dt}\right).
		\label{eq:Lobs}
	\end{equation}
	\noindent In this expression, $(r_{\rm tr}/r_{\rm in})^{-2/3}$ represents the radiation degradation from the inner boundary up to the trapping radius due to adiabatic expansion caused by the advection. 
	The trapping radius as a function of time $r_{\rm tr}(t)$ can be estimated by equating Equations~(\ref{eq:tdif}) and~(\ref{eq:tdyn}).
	Then, by combining it with Equation~(\ref{eq:Lobs}) together with its time derivative it is possible to calculate the exact value of this expression as a function of time.
	In addition, \cite{P20} calculated the scaling relations for the different stages of the transient development that agree with the exact solution, 
	\begin{equation}
		\frac{L_{\rm obs}}{L_*} \propto
		\begin{cases}
			t^{-1/2}, & v_{\rm w}t / r_{\rm in} < 1\\
			t^{-1/6}, & v_{\rm w}t / r_{\rm in} \gg 1\\
			A^{-2/3}, & v_{\rm w}t / r_{\rm in} \gg A.
		\end{cases}
	\end{equation}
	\noindent In the initial phase ($v_{\rm w}t / r_{\rm in} < 1$), the trapping radius is similar to the wind outermost radius so that advection is basically the only transport mechanism of radiation. 
	Then, this does not hold anymore as the wind extension becomes larger than the trapping radius but keeps increasing albeit at a slower pace. 
	Therefore, radiation is advected only up to the trapping radius and then diffuses the rest of its way up to the edge of the wind. 
	This explains the shallower power law. 
	Finally, the system evolves to a time-steady configuration of maximum attenuation.
	
	So far this description only applies to a wind with constant $A$, i.e. constant mass-loss rate and opacity. 
	However, \cite{P20} also explored the case for a time-dependent wind. 
	Let us consider the same problem but now allowing the wind to evolve as a function of time,
	\begin{equation}
		A(t)=A_{\rm max}(1+t/t')^{-\beta},
		\label{eq:mdot}
	\end{equation}
	\noindent where $A_{\rm max}$ is a constant, $t'$ a characteristic timescale for the wind evolution, and $\beta$ the power law of the evolution of the wind properties.
	Including this term into the analytical formalism, the modified scaling relations of the ratio of the observed to injected luminosity are
	
	\begin{equation}
		\frac{L_{\rm obs}}{L_*} \propto
		\begin{cases}
			t^{-1/2}, & t \lesssim r_{\rm in} /v_{\rm w}\\
			t^{-1/6}, &  r_{\rm in} /v_{\rm w}\ \lesssim t \lesssim t'\\
			A(t)^{-2/3}, &  t' \lesssim t \lesssim t'A_{\rm max}^{1/\beta}\\
			1, & t \gtrsim A^{1/\beta}_{\rm max}.
		\end{cases}
		\label{eq:scaling-mdot}
	\end{equation}
	
	\noindent In this case the first two stages of the evolution are identical to the constant wind case as long as the characteristic timescale is long enough to allow both phases. 
	At $t>t'$ the wind starts to become transparent either due to a decrease in its mass-loss rate or opacity. 
	As a result, the trapping radius decreases causing radiation to be less affected by the wind. 
	Then, more radiation manages to go through the wind, which increases the observed luminosity. 
	This stops only once the wind becomes fully transparent and radiation is no longer reprocessed but rather it propagates freely. 
	
\section{Numerical simulations}
\label{sec:numerical}

	\subsection{Equations and method}
	\label{sec:eq_meth}
        We solve the radiation-hydrodynamic equations derived in the mixed-frame formulation following \cite{K07}, where we keep terms up to $\mathcal{O}(u/c)$. 
	    As a closure relation we use the FLD approximation \citep{A73}.
	    Within this approach the radiative flux is written as a function of the radiative energy gradient following Fick’s law. 
	    For further details of the equations and their derivation we refer the reader to the Appendix~\ref{sec:app1}.

    	The set of equations is solved making use of the operator-splitting technique. 
    	Firstly, we use the Godunov method with second-order time evolution already included in JET. 
    	Secondly, we evolve explicitly the sink/source terms due to the radiation coupling, such as radiation pressure and work exerted by radiation. 
    	Finally, there is an implicit integration step to evolve the thermal energy and the radiation energy to simulate the radiation absorption, emission and diffusion processes. 
    	The method for performing this step is derived and detailed in the Appendices~\ref{sec:app1} and~\ref{sec:app2}.
	
	\subsection{Wind ejection and irradiation}
	\label{sec:wind_ej}
	
	    In order to simulate a steady wind  as well as the irradiation of a central source we included extra prescriptions into the code.
	    First, the wind was modelled as a spherically symmetric source term of mass, momentum, and energy according to the free-wind analytical solution.
	    The parameters to specify the wind properties are the mass-loss rate $\dot{M}$, terminal velocity $v_{\rm w}$ (assumed to be constant), the pressure $p_0$ (or temperature $T_0$) of the wind at an arbitrary radius $r_0$. 
    	The injection of mass, momentum, total energy, and radiation energy (assumed to be negligible so set to the floor value) is done by explicit source terms at the innermost radial boundary.
	
	    Second, the irradiation is included as an inner boundary condition, which is considered in the implicit diffusion step. 
	    The source is specified by its luminosity $L_*$, and the opening angle $\theta_{\rm op}$, which it illuminates. 
    	If the wind considered is optically thick (as is typical in our case)
    	it is necessary to consider an extra decay factor in the irradiation boundary condition due to adiabatic energy degradation in order to compensate for the the mesh displacement.
    	If the trapping radius is larger than the inner radius of the domain, we multiply the irradiation radiative flux by a factor $(r_{\rm in,mesh}/r_{\rm in})^{-2/3}$. 
    	On the contrary, if the trapping radius is smaller than the inner radius of the domain, we use a factor $(r_{\rm tr}/r_{\rm in})^{-2/3}$.
    	
        \begin{table}
            \centering
            \begin{tabular}{l|c|c|c|c|c|c}
            \hline
            Name    &   $\log(A)$   & $\theta_{\rm op}$ &   $(v_{\rm w}t'/r_{\rm in},\beta)$  &   $v_{\rm w}t_{\rm f}/r_{\rm in}$ & $N_r\times N_{\theta}$\\
            \hline
            \hline
            A20        &   2.0 & $90\degr$ &    --  &  $10^{5.5}$ & $512\times256$\\
            A25        &   2.5 & $90\degr$ &    --  & $10^{5.5}$ & $512\times256$\\
            A30        &   3.0 & $90\degr$ &    --  & $10^{5.5}$ & $512\times256$\\
            \hline
            A20-op15   &   2.0 & $15\degr$ &    --  & $10^{5.5}$ & $512\times256$\\
            A20-op45   &   2.0 & $45\degr$ &    --  & $10^{5.5}$ & $512\times256$\\
            \hline
            A40-t01    &   4.0 & $90\degr$ &    $(10^{-2},5/3)$  & $10^4$ & $512\times256$\\
            A40-t1     &   4.0 & $90\degr$ &    $(1,5/3)$  & $10^6$ & $512\times256$\\
            A40-t2    &   4.0 & $90\degr$ &    $(10^2,5/3)$  & $10^{9}$ & $512\times256$\\
            A40-t4    &   4.0 & $90\degr$ &    $(10^4,5/3)$  & $10^9$ & $512\times256$\\
            \hline
            \end{tabular}
            \caption{Simulated models and their parameters. 
            Column 1: the name of the model. 
            Column 2: value of the dimensionless parameter $A$ (see Eq.~[\ref{eq:A}]). 
            Column 3: irradiation opening angle. 
            Column 4: characteristic timescale and power law for the wind evolution (see Eq.~[\ref{eq:mdot}]). 
            Column 5: simulation final time. 
            Column 6: initial resolution in the radial $N_r$ and polar $N_{\theta}$ directions.}
            \label{tab:models}
        \end{table}	
	
	\subsection{Setup and parameters}
	\label{sec:setup}
	
		We consider a two-dimensional domain in spherical coordinates $(r,\theta)$, i.e. we assume azimuthal symmetry. 
		The initial domain lengths are $r_{\rm in}\le r\le 10r_{\rm in}$ and $0\le \theta \le 90\degr$ in the radial and polar directions, respectively. 
		The setup is such that the radial dimension spans up to two orders of magnitude at any given time. 
		The initial grid is made out of $512$ logarithmically spaced cells in the radial direction, and $256$ linearly spaced cells in the polar direction.
		It is important to remark that the number of radial cells may change through the simulation as the inner and outer boundaries can move at different speeds, though most of the simulation time the number of radial cells is $\sim$1000. 
		The outer boundary conditions were set to outflow or zero-gradient conditions for all variables except for radiation energy density, which was set to free-stream condition so that the radiation can escape freely. 
		We use a HLL Riemann solver for the hydrodynamics hyperbolic step. 
		The radiation implicit step was solved with a Stabilised Bi-Congujate Gradient linear solver \citep{V92} together with the BoomerAMG preconditioner \citep{Y02} from the \textit{hypre} library \citep{F06}. 
		We use them with their default parameters except for the relative convergence tolerance, which was set to $10^{-10}$.
		These choices guaranteed rapid convergence (in less than 10 iterations) as well as stability in the iterative solver according to our empirical tests.
		The domain was initialised at rest $\bf{u}=\bf{0}$ with density $\rho = (r_{\rm in}/r)\times10^{-12}\,\rm g\ cm^{-3}$, gas temperature $T=(r_{\rm in}/r)\times10^3\rm\ K$, and radiation temperature $T_{\rm r}=10\rm\ K$. 
		These initial conditions allow that the wind and radiation can propagate freely even at later times when the outer edge of the domain becomes significantly larger. 
		All models consider $v_\text{w}=10^8\rm\ cm\ s^{-1}$, $r_\text{in}=10^{10}\rm\ cm$, and $\dot{M}=1.75A\times10^{-4}\rm\ M_{\odot}\ yr^{-1}$. 
		The central luminosity is constant and set to $L_*=7.125\times10^{36}\rm\ erg\ s^{-1}$, which is the value corresponding to a spherical source of radius $r_{\rm in}$ and $T_{\rm eff}=10^5\rm\ K$. 
		Notice that with these choices the ratio of the injected radiation to wind momenta is $L_*/(\dot{M}v_{\rm w}c)<2.14\times10^{-6}$. 
		Thus the radiation has no impact on the wind dynamics. 
		Here we focus only on models within this regime.
	    We assume that the opacity is dominated by electron scattering, i.e. $k_{\rm s}=0.34\rm\ cm^2\ g^{-1}$, which is reasonable for the range in density and temperature considered. 
	    Finally, we set $k_\text{P} = 0$ as the opacity is scattering dominated, and also so that we can make direct comparisons with previous analytical estimates.
		 
		Our models can be divided into three categories: isotropic irradiation with constant mass-loss rate, anisotropic irradiation with constant mass-loss rate, and isotropic irradiation with evolving mass-loss rate. 
		The first group considers three models specified by the dimensionless parameter $A=10^2$, $10^{2.5}$, and $10^3$. 
		The second includes two models that use $A=10^2$ but are specified by the irradiation opening angle used $\theta_{\rm op}=15\degr$ and $45\degr$. 
		All the aforementioned models were simulated up to a time of $v_{\rm w}t/r_{\rm in}=10^{5.5}$. 
		Finally, we ran a set of simulations for the case of a wind evolving according to Equation~(\ref{eq:mdot}).
		Motivated by the observed features in tidal-disruption events, we fixed $A_{\rm max}=10^4$, $\beta=5/3$, and explored four values for the characteristic timescale $v_{\rm w}t'/r_{\rm in}=10^{-2}$, $1$, $10^2$, and $10^4$. 
		In order to observe all the regimes of the expected evolution we simulated these models up to a time of $v_{\rm w}t/r_{\rm in}=10^9$. 
		Table~\ref{tab:models} summarises all or models  and their parameters.
		
	    \begin{figure*}
	        \centering
	        \includegraphics[width=0.33\textwidth]{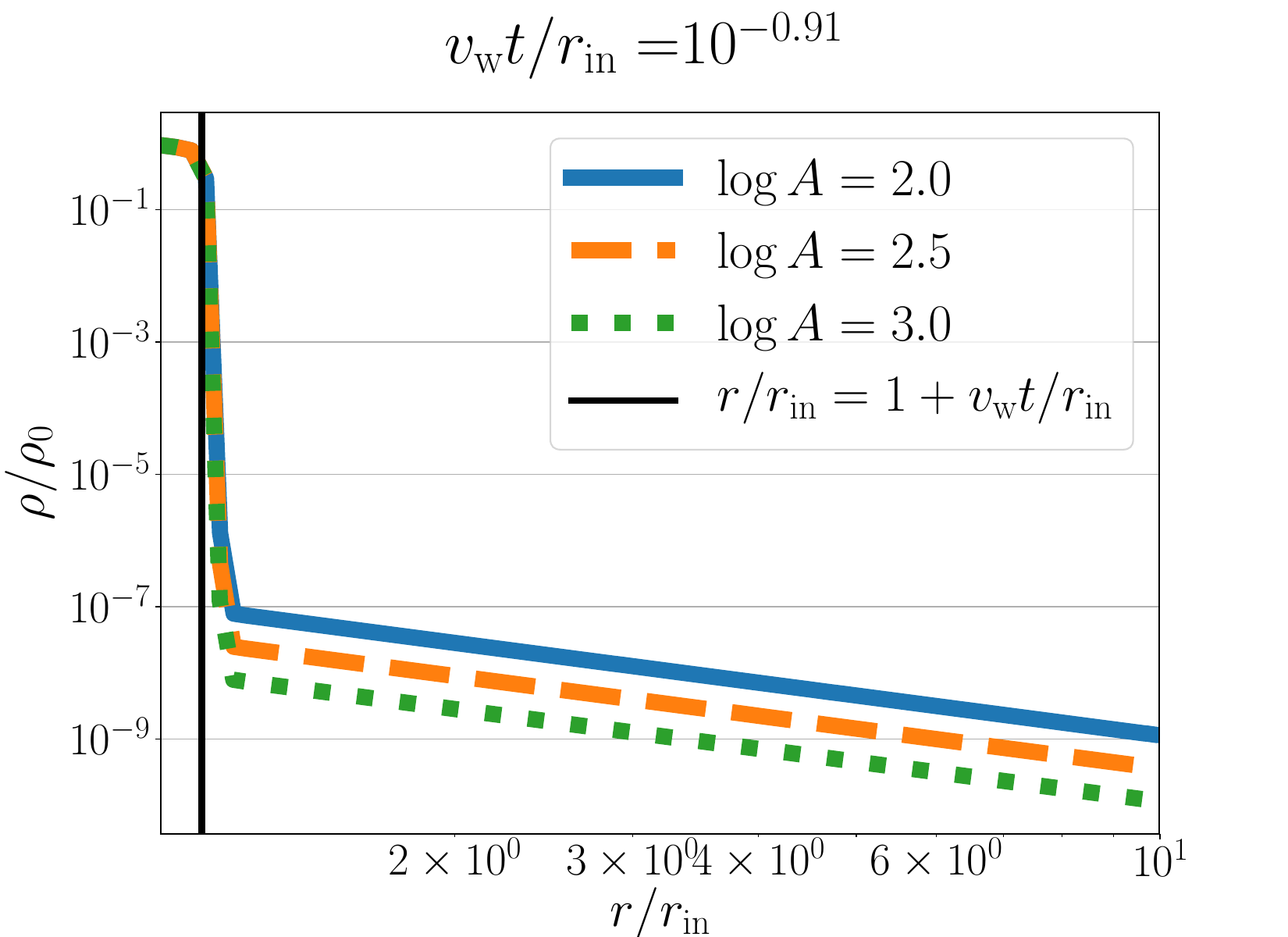}
	        \includegraphics[width=0.33\textwidth]{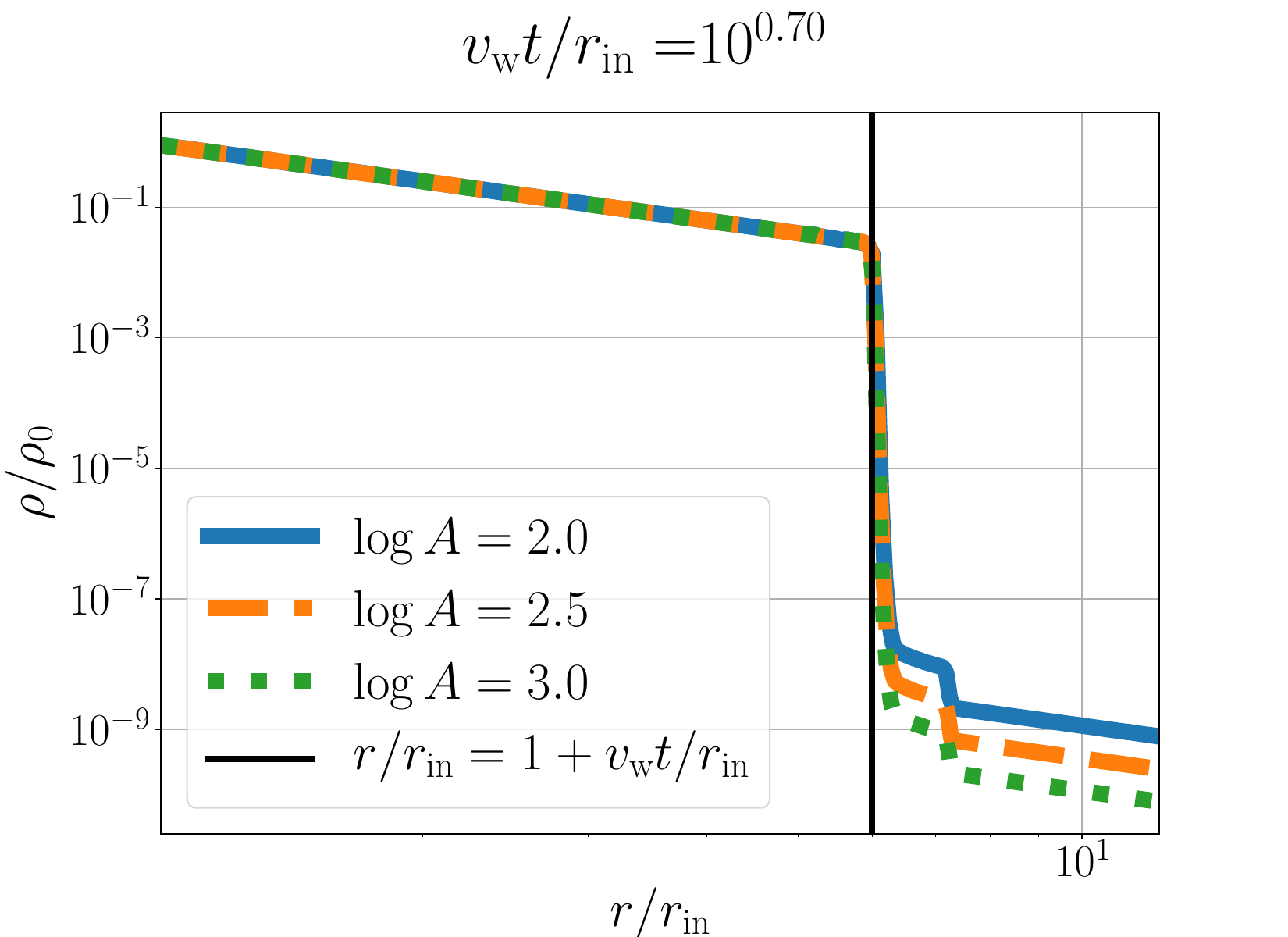}
	        \includegraphics[width=0.33\textwidth]{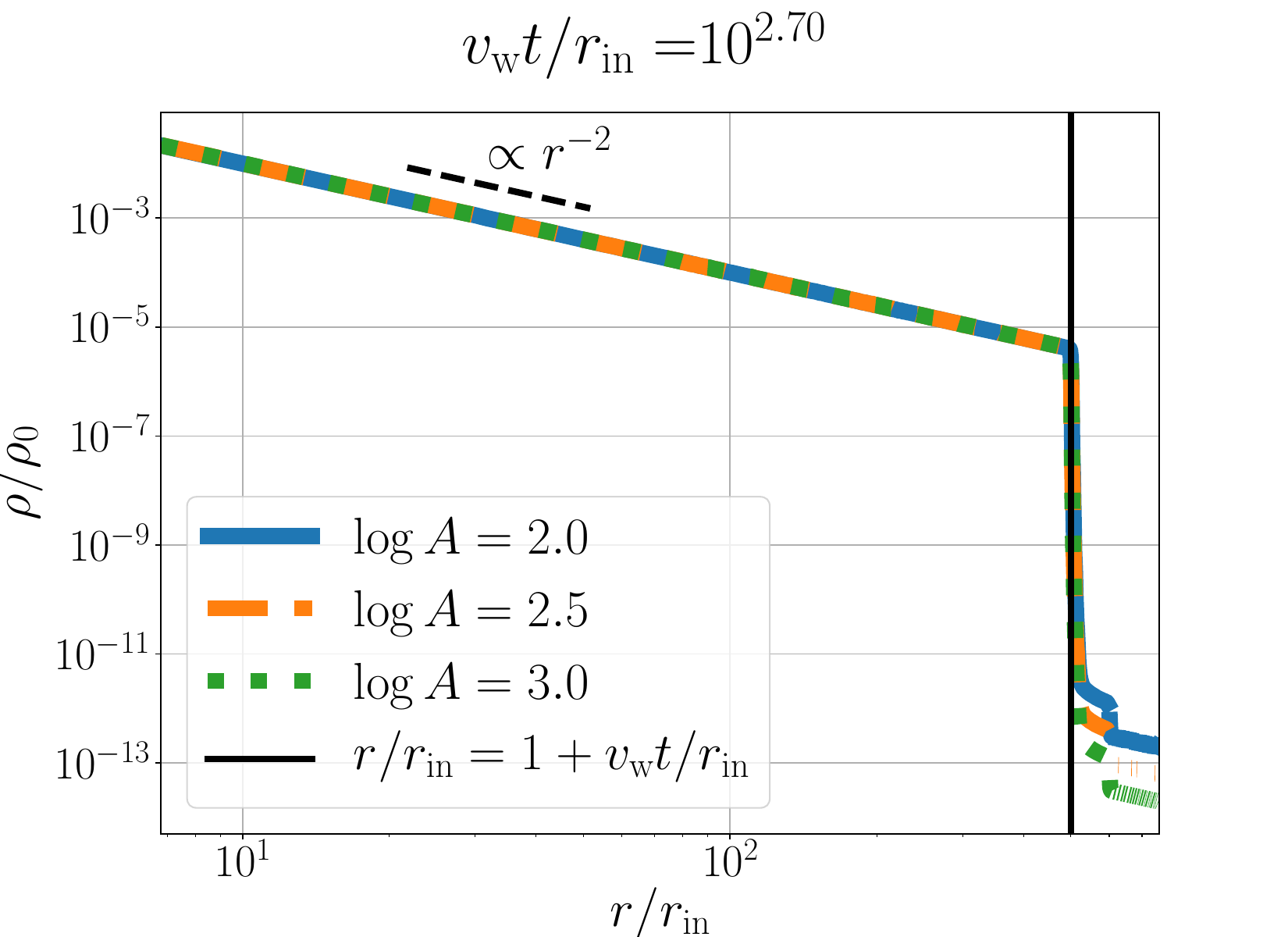}
	        \includegraphics[width=0.33\textwidth]{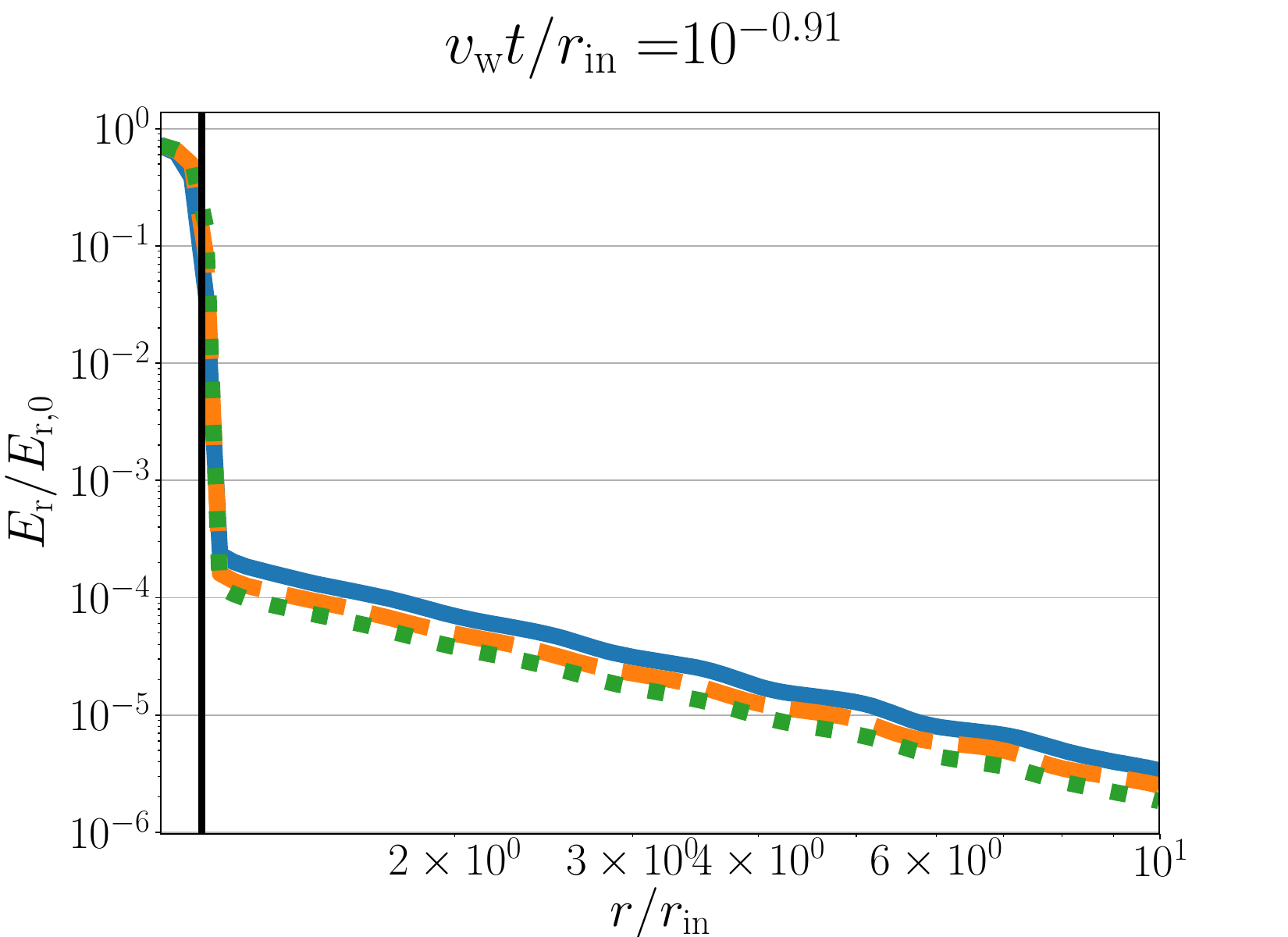}
	        \includegraphics[width=0.33\textwidth]{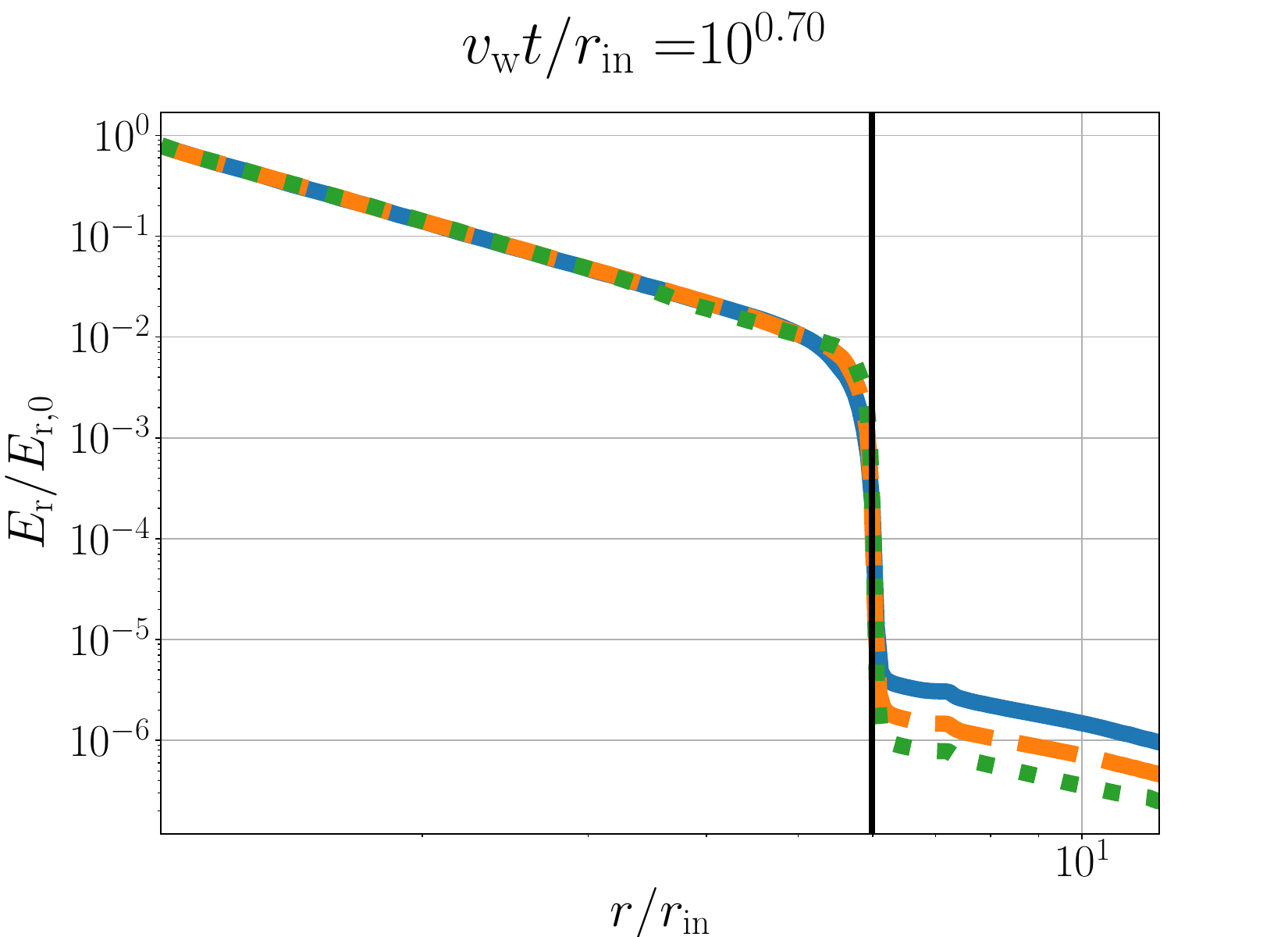} 
	        \includegraphics[width=0.33\textwidth]{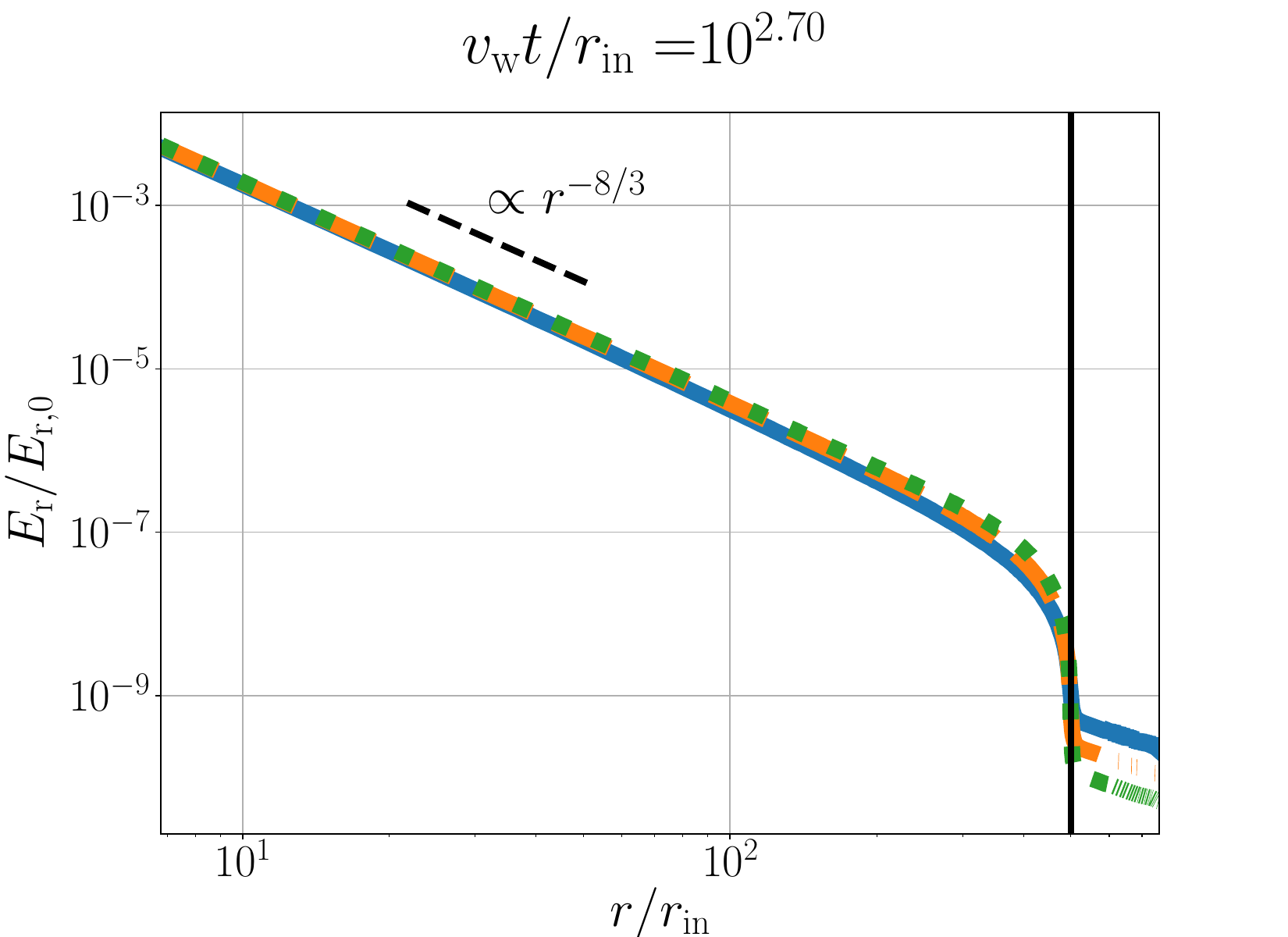} 
	        \includegraphics[width=0.33\textwidth]{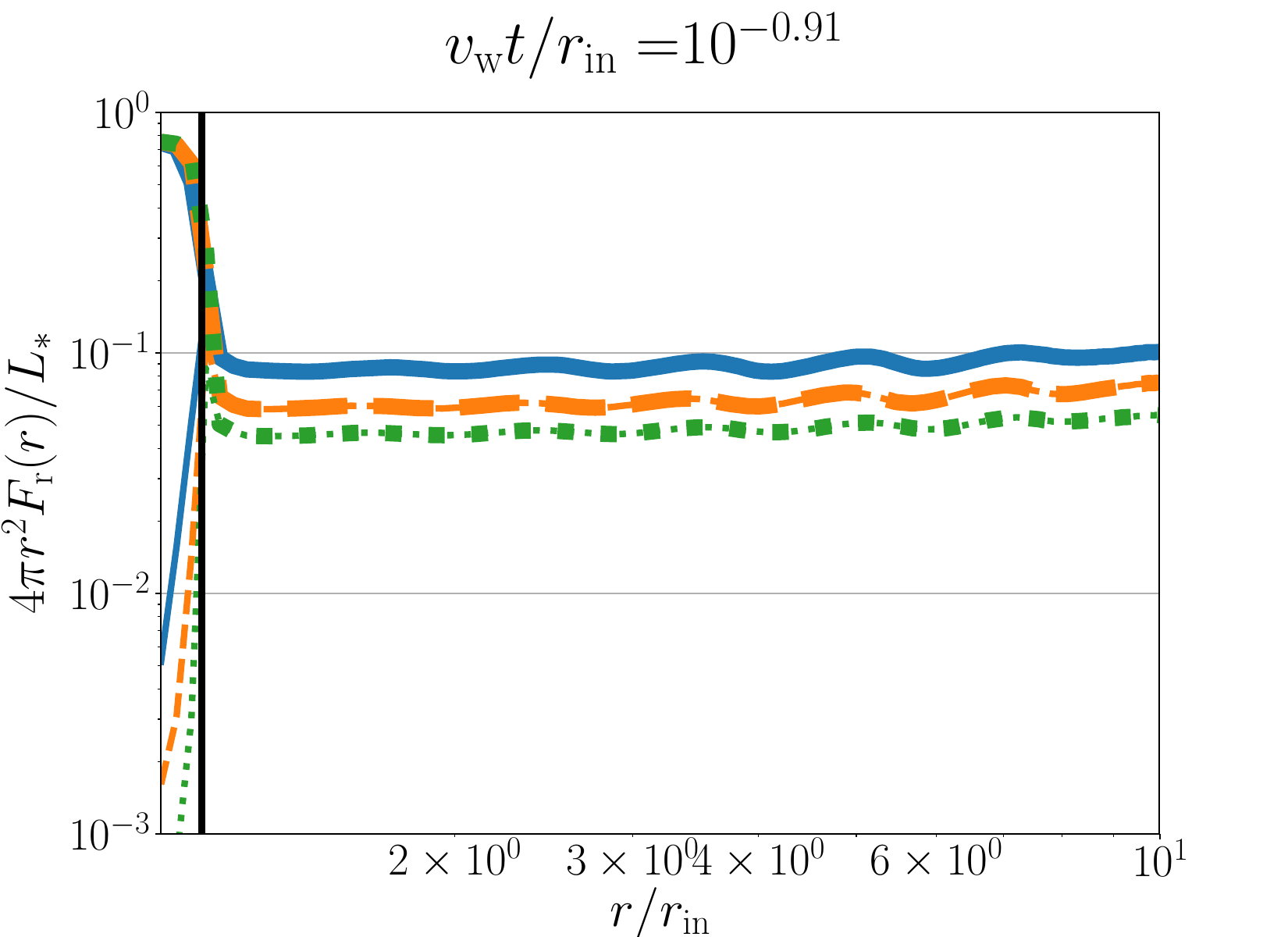}
	        \includegraphics[width=0.33\textwidth]{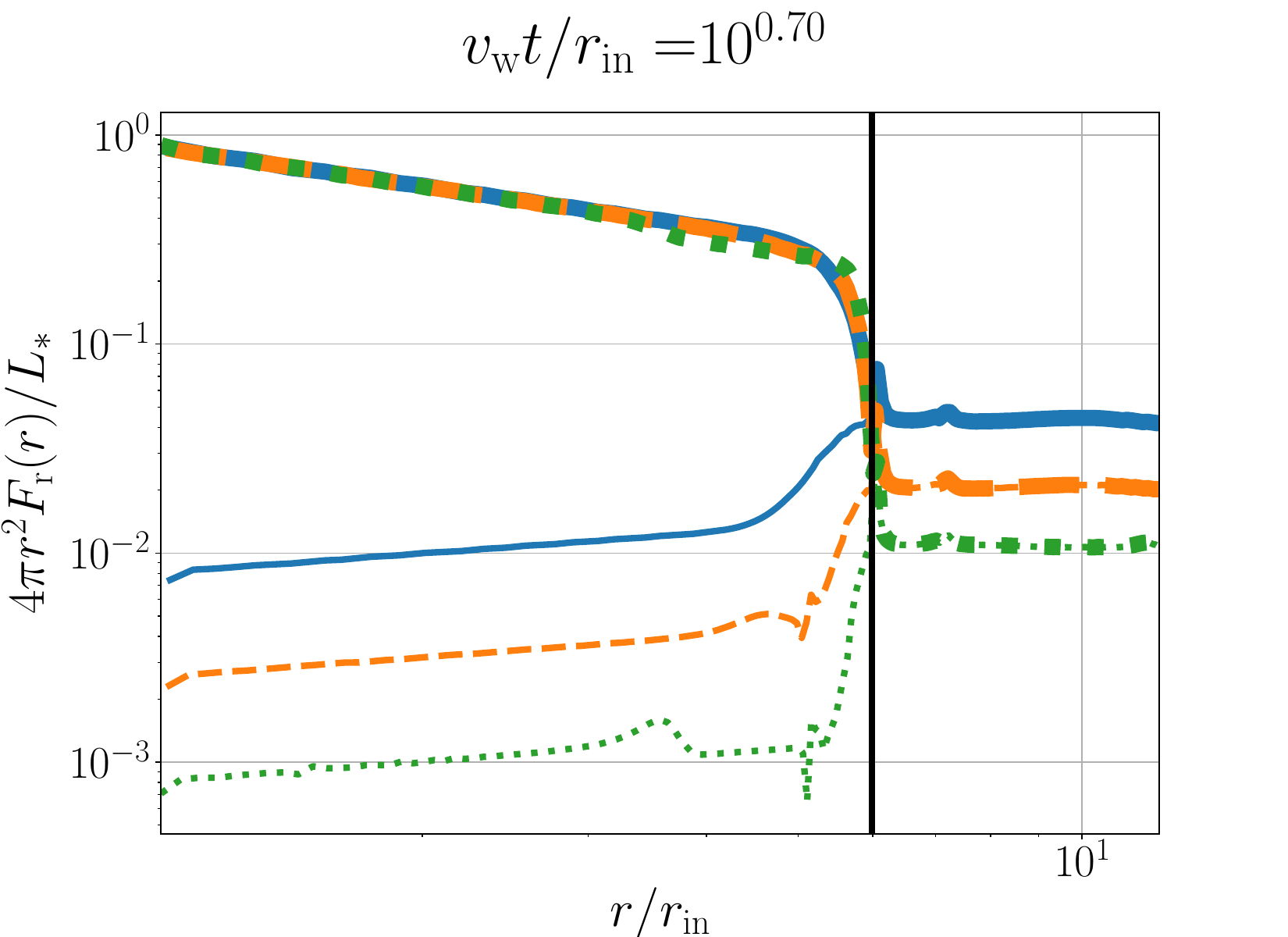} 
	        \includegraphics[width=0.33\textwidth]{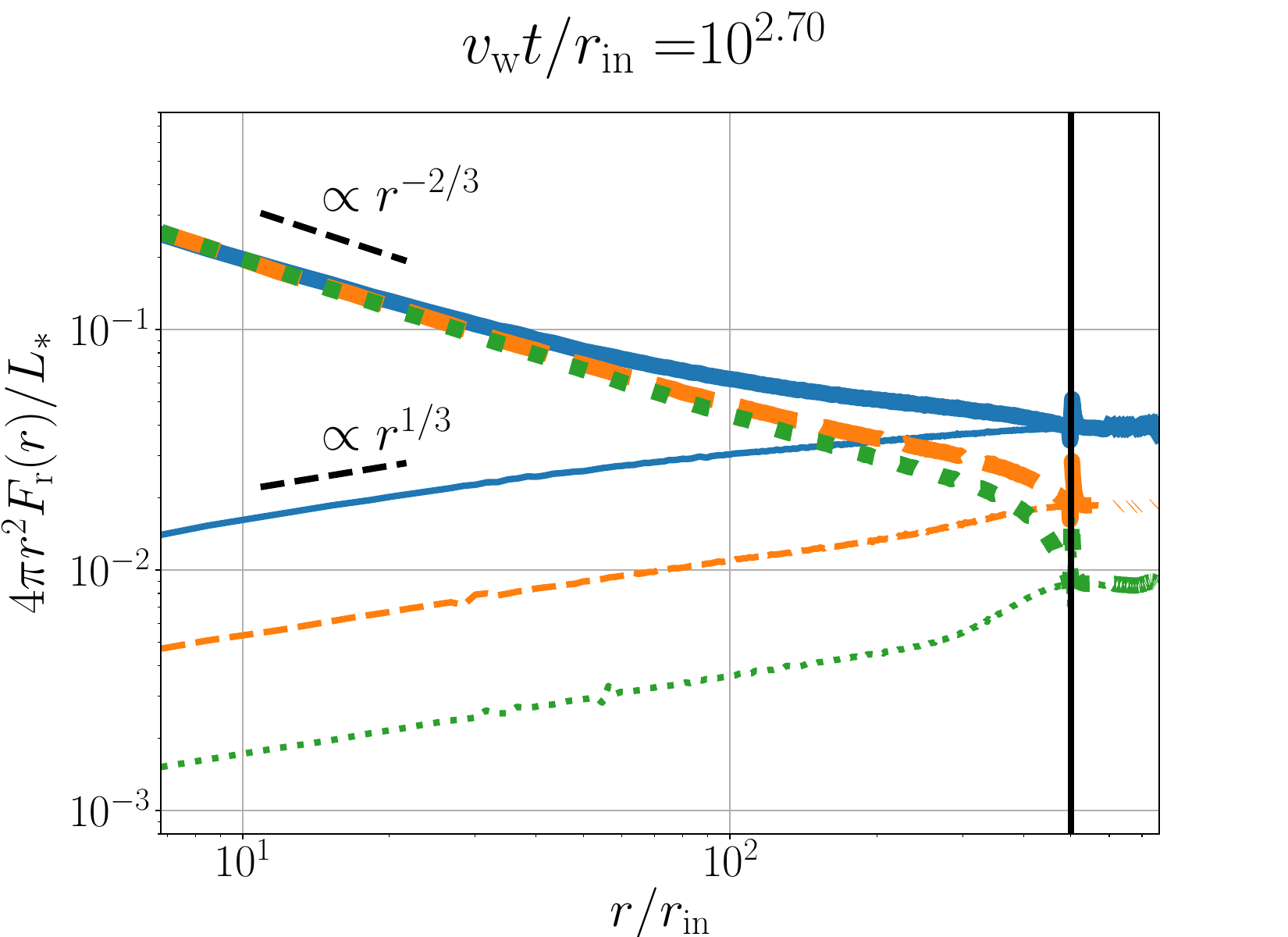} 
	        \caption{
	        Radial profiles of density (top row), radiation energy density (middle row), and spherically-averaged luminosity (bottom row) for models of isotropic irradiation at times $v_{\rm w}t/r_{\rm in}=10^{-0.91}$ (left column), $10^{0.7}$ (middle column), and $10^{2.7}$ (right column). 
	        Each panel shows models with $A=10^{2}$ (solid blue line), $A=10^{2.5}$ (dashed orange line), and $A=10^3$ (dotted green line).  
	        All variables are scaled to their values at the injection radius $r_{\rm in}$.
	        The average luminosity radial profiles (bottom row) include thinner lines, which represent solely the diffusive component of the luminosity. 
	        The solid black vertical line represents the outermost extension of the wind $r_{\rm w}$. 
	        The short dashed black lines are the expected analytical scaling relation for each variable.
	        }
	        \label{fig:snapshot}
	    \end{figure*}
	
	    \begin{figure}
	        \centering
	        \includegraphics[width=0.495\textwidth]{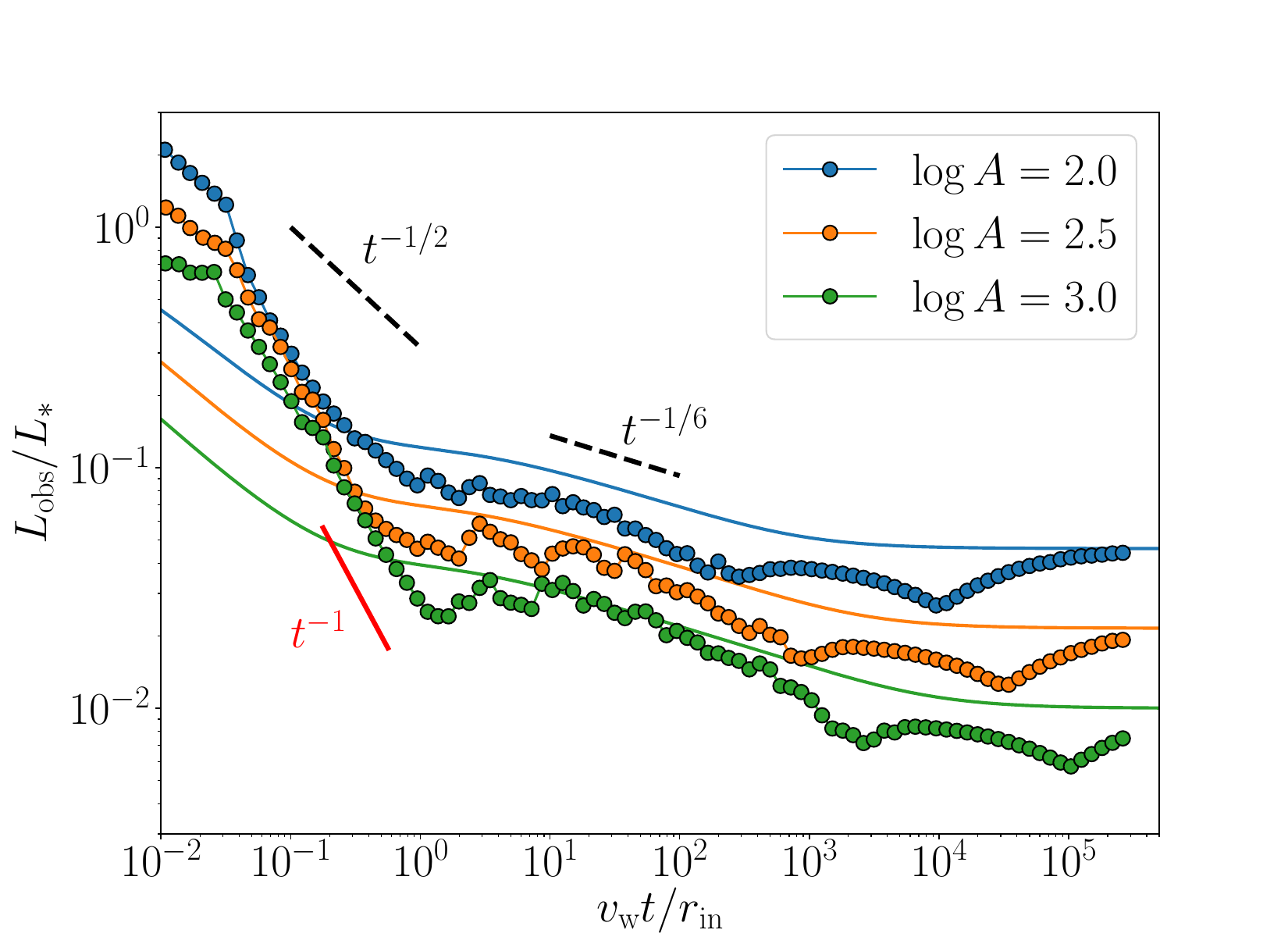} 
	        \caption{
	        Analytical (solid lines) and simulated (connected dots) light curves of isotropically irradiated wind-reprocessed transients with $A=10^{2}$ (blue), $10^{2.5}$ (orange), and $10^3$ (green). 
	        $L_\text{obs}$ was rescaled by the injected luminosity $L_*$ at the inner radius $r_{\rm in}$. 
	        }
	        \label{fig:comparison}
	    \end{figure}
	    
	    \begin{figure}
	        \centering
	        \includegraphics[width=0.495\textwidth]{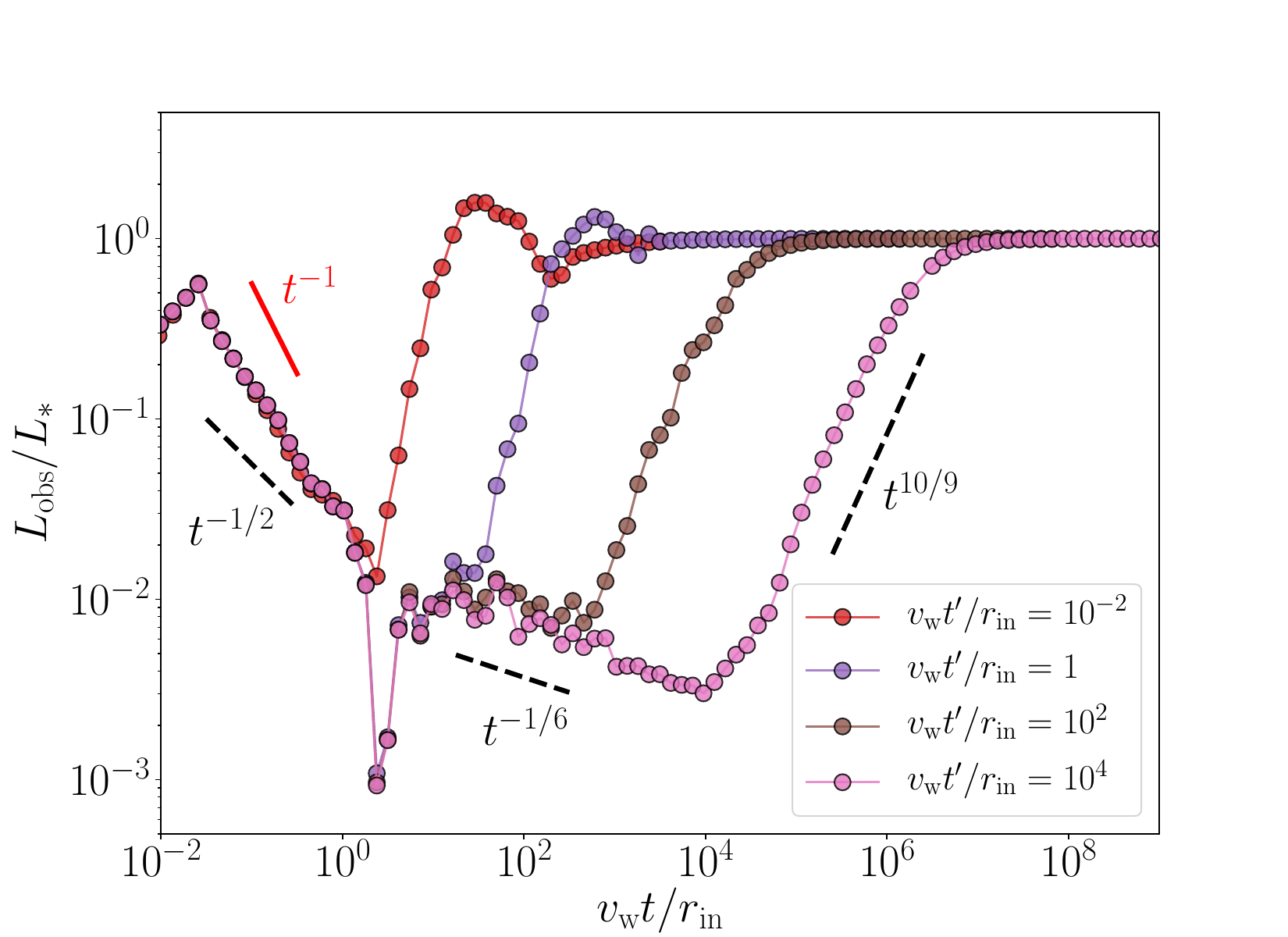} 
	        \caption{
	        Simulated light curves for isotropically irradiated wind-reprocessed transient with an evolving mass-loss rate with $A_{\rm max}=10^4$, $\beta=5/3$, and characteristic timescales $v_{\rm w}t'/r_{\rm in}=10^{-2}$, $1$, $10^2$, and $10^4$ shown as connected points in red, purple, brown, and pink, respectively. 
	        Notice that the $x$ axis spans over $\sim11$ orders of magnitudes.
	        }
	        \label{fig:mdot}
	    \end{figure}
	    
	    \begin{figure*}
	    \centering
	        \includegraphics[width=0.4\textwidth]{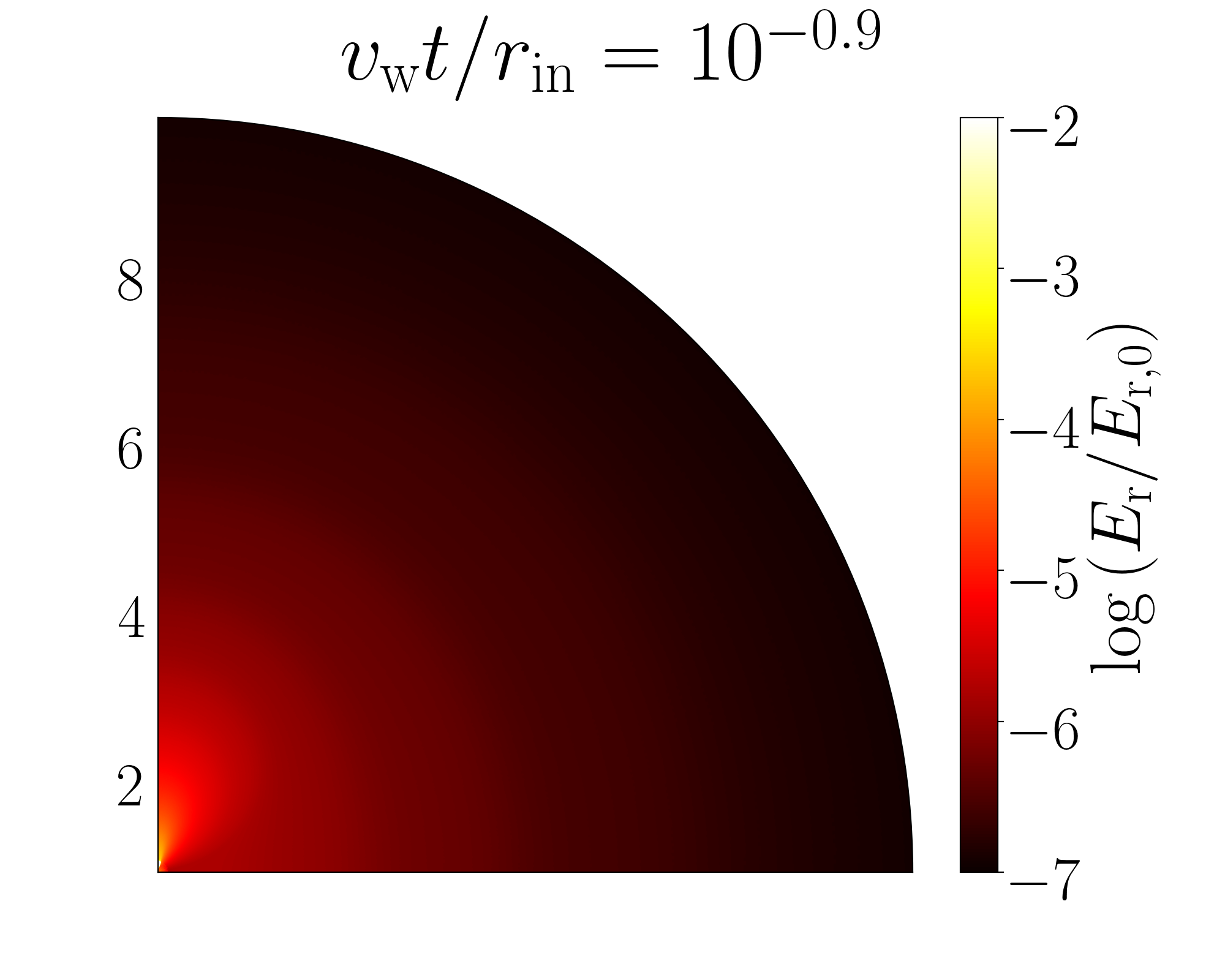}
	        \includegraphics[width=0.4\textwidth]{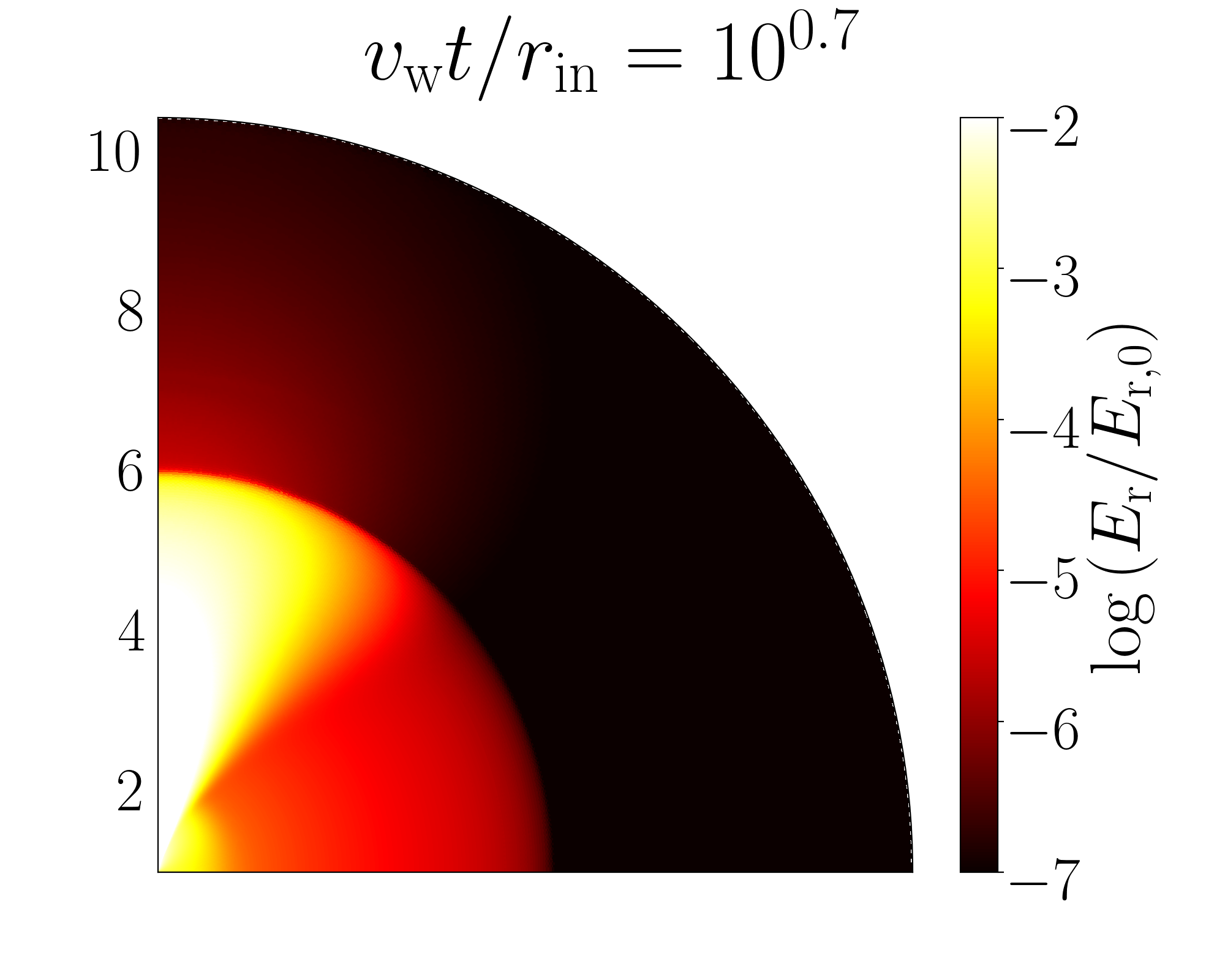}
	        \includegraphics[width=0.4\textwidth]{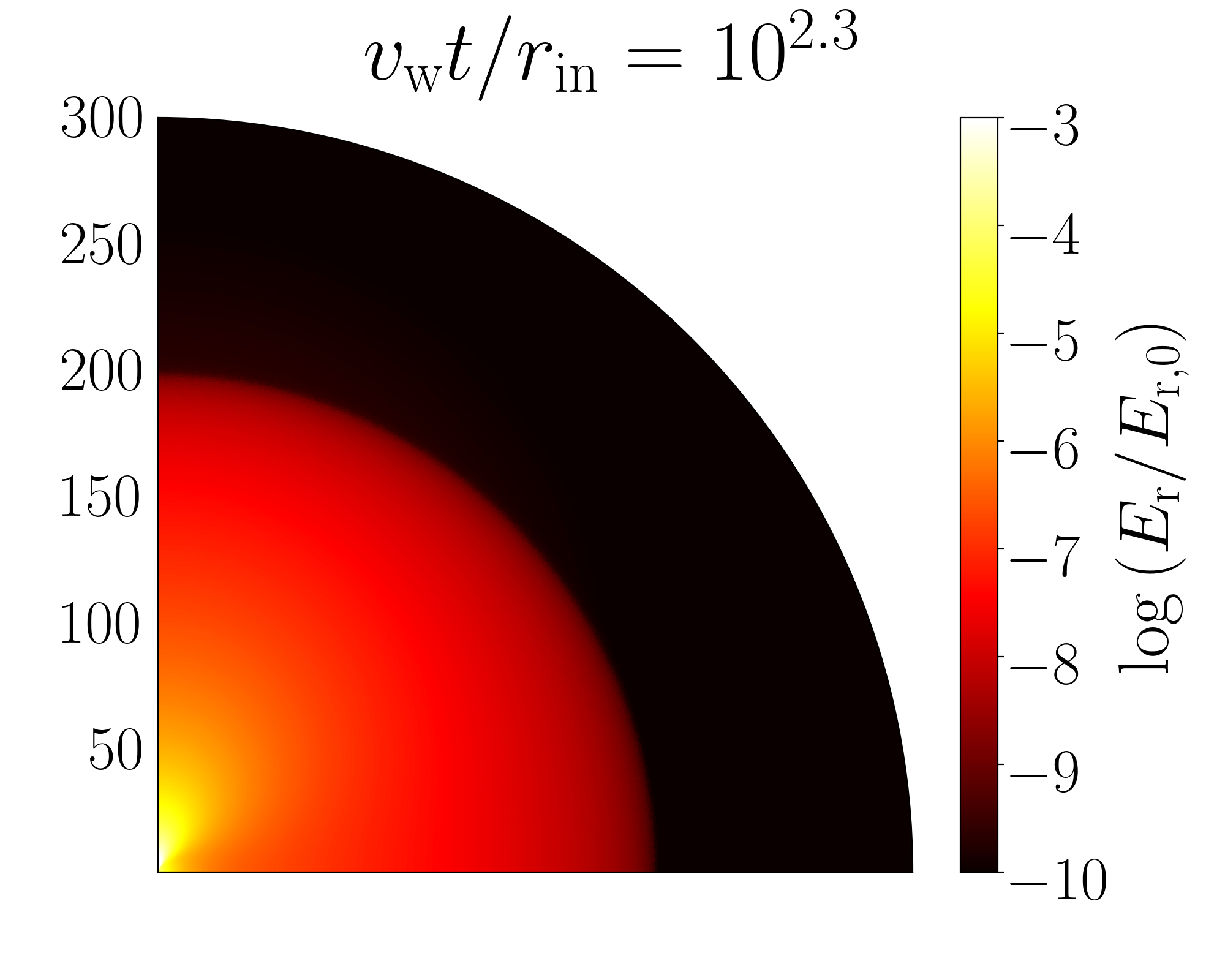}
	        \includegraphics[width=0.4\textwidth]{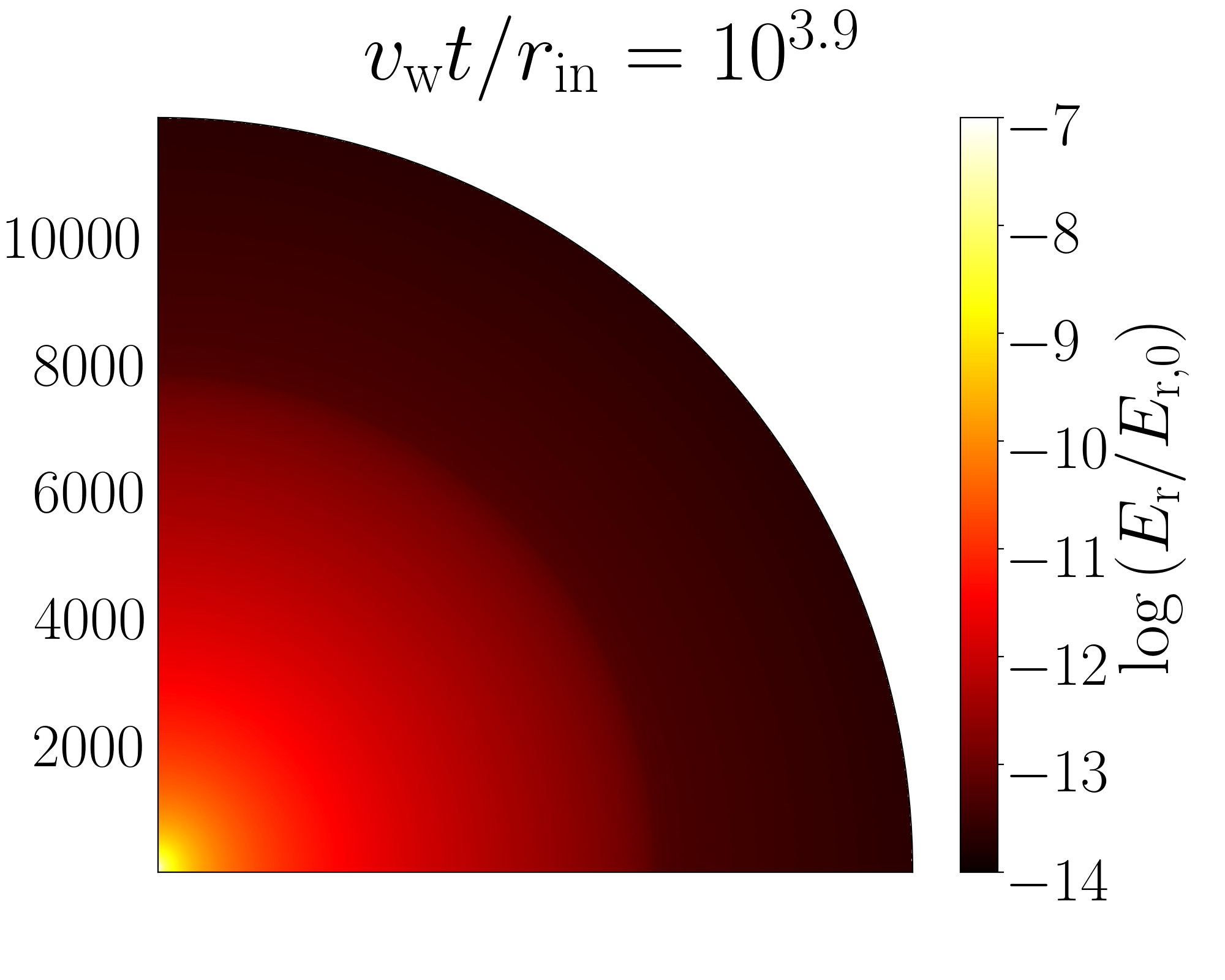}
	        \caption{Radiation energy density maps for model with $A=10^2$ and $\theta_{\rm op}=15\degr$, i.e. the irradiation only is injected for $\theta\leq\theta_{\rm op}$. Panels show snapshots at times $v_{\rm w}t/r_{\rm in}=10^{-0.9}$, $10^{0.7}$, $10^{2.3}$, and $10^{3.9}$ (from top left to bottom right).
	        Notice the changes in the physical scale and dynamic range due to the radial motion of the grid.}
	        \label{fig:maps}
	    \end{figure*}
	    
	    \begin{figure*}
	   		\centering
	   		\includegraphics[width=0.495\textwidth]{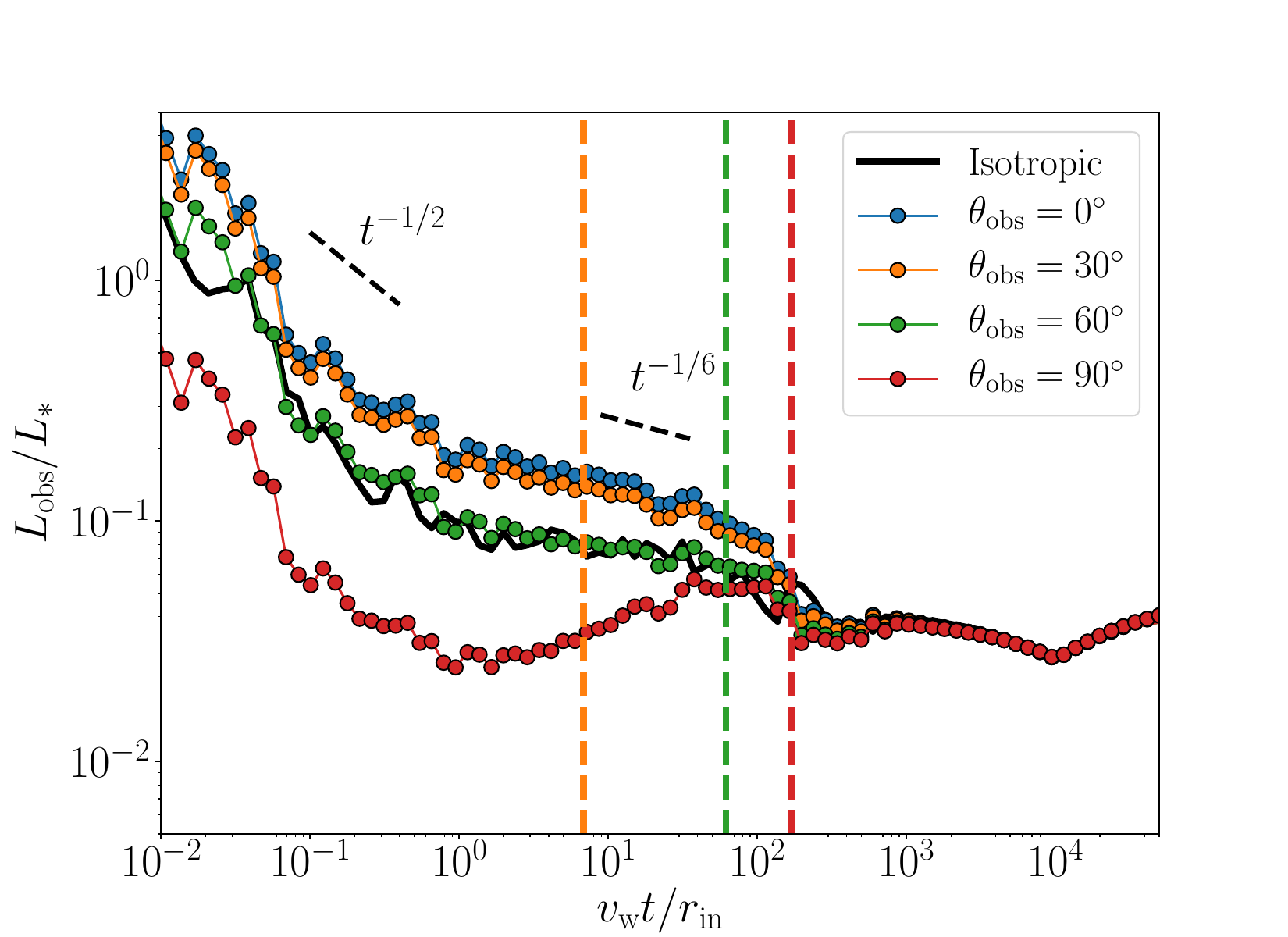}
	   		\includegraphics[width=0.495\textwidth]{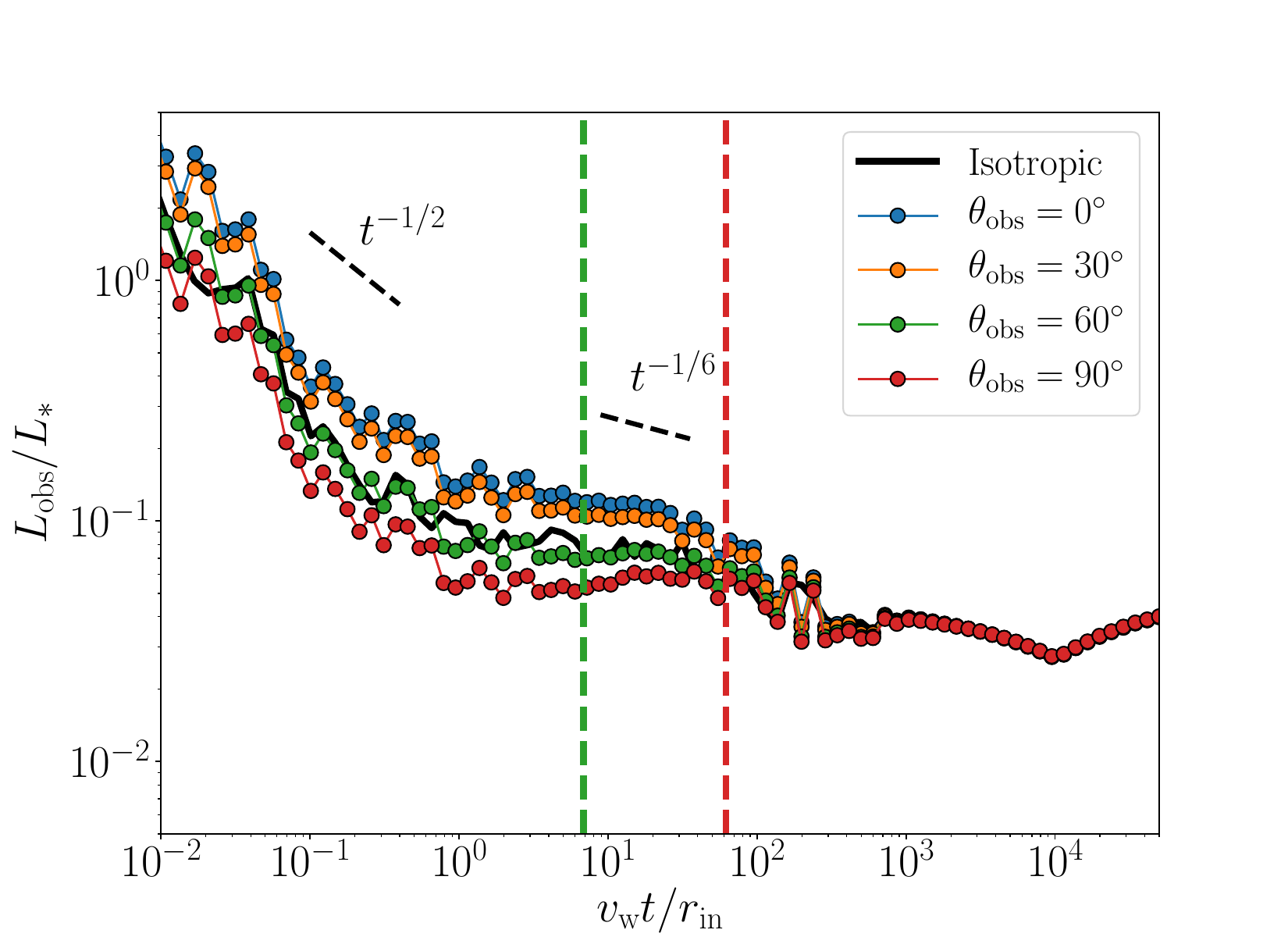} 
	   		\caption{
	    			Simulated light curves for anisotropically irradiated
	    			wind-reprocessed transient with a constant wind whose $A=10^{2}$ with irradiation opening angle of $\theta_{\rm op}=15\degr$ (left panel), and $\theta_{\rm op}=45\degr$ (right panel).
	    			Light curves are shown for different lines of sight for every model: $\theta_\text{obs}=0\degr$ (blue dots), $30\degr$ (orange dots), $60\degr$ (green dots), and $90\degr$ (red dots). 
	    			The simulated light curve of the isotropic irradiation case with $A=10^2$ is shown as a solid black line.  
	    			The vertical dashed lines represent the circular diffusion timescale for a given line of sight. 
	    			Notice that it is has been scaled by the injected luminosity at the inner radius $r_{\rm in}$. 
	    		}
	   		\label{fig:asym}
		\end{figure*}
		
\section{Results}
\label{sec:results}

	\subsection{Isotropic irradiation with constant wind}
	\label{sec:iso}
	
		Figure~\ref{fig:snapshot} shows radial profiles of the evolution of the isotropic irradiation case with a constant wind. 
		The top, middle, and bottom rows display density, radiation energy density, and average luminosity, respectively. 
		In addition, the left, central, and right columns show equal simulation times: $v_{\rm w}t/r_{\rm in}=10^{-0.91}$, $10^{0.7}$, $10^{2.7}$, respectively. 
		Each panel presents models with $A=10^2$ (solid blue line), $10^{2.5}$ (dashed orange line), and $10^3$ (dotted green line). 
		We have marked the outermost extension of the wind with a solid black vertical line. 
		As a reference, we included short dashed black lines with the slopes expected from analytical estimations.
		
		In the density radial profiles (top row), we see the wind expansion shapes the expected free-wind profile $\propto r^{-2}$. 
		Through the entire evolution it is possible to observe a big jump in density (about a factor $10^8$) at the outermost extension of the wind due to the presence of a shock connecting the ejected wind with low-density medium. 
		At later times (top central and right panels) it can be seen how the low-density material is compressed ahead of the wind though it is density is much smaller than the wind's.
		
		The radiation energy density radial profiles (middle row) show an analogous time evolution compared to the density profiles but following a $\propto r^{-8/3}$ decay instead.  
		This is the result of the adiabatic expansion experienced by the radiation as it is advected together with the wind. 
		Reaching the outer extension of the wind there is a steep transition due to the drastic change from an optically thick to an optically thin medium. 
		Once the radiation reaches this point it can escape freely. 
		The panels of the bottom row show the averaged luminosity as a function of radial distance. 
		This luminosity is calculated as the averaged radiative flux multiplied by the surface area of a sphere with radius $r$, i.e. $4\pi r^2F_\text{r}(r)$.
		This quantity has been rescaled by the luminosity of the central source $L_*$. 
		Here it is important to bear in mind that in this case there are two mechanisms for radiation transport: advection and diffusion. 
		Thus, the radiative flux can be expressed as the sum of both of its components. 
		In these panels, the thicker lines represent the total radiative flux while the thinner ones stand solely for the diffusion component. 	
		Initially, the diffusion is not an efficient mechanism for radiative transport. 
		This can be seen as, in general, it contributes a small fraction of the total luminosity (thick lines) but its importance increases closer to the edge of the wind. 
		Nevertheless, at later times (bottom central and right panels) diffusion becomes progressively more relevant within the wind. 
		Notice that in the case model $A=10^2$ the diffusion component is as important as the advection component close to the edge of the wind. 
		This is occurs as the wind becomes more dilluted the radiation starts to decouple from the wind, so that it can diffuse through the medium. 
		The profiles are also reproduced very well by the analytical scalings. 
		The total luminosity decays with the expected $\propto r^{-2/3}$ slope while the diffusion component increases radially following the power law $r^{1/3}$ as a result of applying the FLD approximation on the radiation energy density profile. 
		
		Now, we proceed to calculate the numerical light curves based on our simulations. 
		This was done by identifying the location of the photosphere, i.e. the position along radial rays where  optical depth is $\tau=2/3$. 
		During the initial evolution the location of the photosphere is within the shell of compressed material slightly ahead of the outermost extension of the wind. 
		However, at later times the photosphere migrates inside the wind as the density decreases. 
		We measured the radiative flux at such locations through all snapshots of all models. 
		Then, we computed the observed averaged luminosity $L_{\rm obs} = 4\pi r^2F_\text{r}|_{\tau=2/3}$ as a function of dimensionless time. 
		The results are shown in Figure~\ref{fig:comparison} for all three models: $A=10^2$ (blue), $10^{2.5}$ (orange), and $10^3$ (green). 
		The dots and lines represent the numerical results and the analytical estimates, respectively. 
		Each dot corresponds to a single snapshot of each of the simulation.
		The latter were calculated by using Equation~(\ref{eq:Lobs}).
		Also, we show the expected power laws in the different regimes as dashed lines for guiding the eye. 
		These results show that overall the numerical simulations reproduce the trends predicted by the analytical analysis. 
		Nevertheless, there are some differences in certain phases of the evolution. 
		At early times ($v_{\rm w}t/r_{\rm in}<1$) the numerical simulations show higher observed luminosity compared to the theoretical prediction. 
		A discrepancy at this stage is not unexpected as \cite{P20} warned that some features might differ as the wind needs a timescale $\sim v_{\rm w}/r_{\rm in}$ to develop as well as for the density profile to be well sampled. 
		In addition, this phase shows a much steeper decay ($\propto t^{-1}$) compared to the analytical prediction ($\propto t^{-1/2}$), which seems to be clearer for models with larger $A$. 
		We investigated this behaviour further by increasing the resolution of the simulation, or by launching the wind for some time in advance before injecting the radiation. 
		However, the same decay feature was observed. 
		A potential explanation for this behaviour could be the assumption of the analytical model of neglecting diffusion within the wind during this phase. 
		Notice that this might not be appropriate at such early times as its contribution could be of about $10\%$ of the total luminosity close to the edge of the wind (see bottom left panel of Figure~\ref{fig:snapshot}). 
		In the next phase ($v_{\rm w}t/r_{\rm in}\gtrsim1$), the maximum extension of the wind $r_{\rm w}$ starts to differ from the trapping radius $r_{\rm tr}$. 
		As a result, the luminosity decay follows a shallower time power-law than in the previous phase. 
		At this point, the light curves match relatively well the analytical profiles. 
		Although there are deviations from the theoretical predictions, they are very small ($\lesssim0.2~\rm dex$). 
		Notice that these differences are more significant in models with smaller $A$, i.e. less dense winds. 
		This feature could also be caused by the contribution of diffusion to the radiation transport. 
		In the bottom central panel of Figure~\ref{fig:snapshot}, the model $A=10^2$ shows that diffusion is much more relevant compared to the other models, and it increases towards the outermost edge of the wind. 
		Thus, this could cause extra attenuation that the analytical model ignores as it assumes that the radiation escapes as soon as it gets to the trapping radius. 
		Following the evolution, as $r_{\rm tr}$ reaches its asymptotic constant value, the radiation reaches its maximum attenuation. 
		The change of slope observed in the simulated light curves is caused by the fact that $r_{\rm tr}$ becomes smaller than the radius at which radiation is injected, which occurs because our inner boundary is continuously expanding with the rest of the mesh.  
		Besides the initial phase that shows a steeper power-law decay, the spherically-symmetric wind-reprocessed transient simulations agree relatively well with the analytical predictions. 
		
	\subsection{Isotropic irradiation with evolving wind}
	
	    Figure~\ref{fig:mdot} shows the simulated light curves for the case of an evolving wind according to Equation~(\ref{eq:mdot}) using the parameters $A_{\rm max}=10^4$ and $\beta=5/3$. 
	    We present four simulations, each one with a different characteristic timescale $v_{\rm w}t'/r_{\rm in}=10^{-2}$, $1$, $10^2$, and $10^4$. 
	    These are represented as red, purple, brown, and pink connected dots, respectively. 
	    Each dot corresponds to a single snapshot of each of the simulation. 
	    As a reference, we show the analytical scaling relations from Equation~(\ref{eq:scaling-mdot}) as dashed black lines.
	    Notice that these simulations cover up to 11 orders of magnitudes temporarily. 
	    
	    In the initial phase of the evolution ($v_{\rm w}t/r_{\rm in}<1$), we again see a steeper decay ($\propto t^{-1}$) compared to the analytical prediction ($\propto t^{-1/2}$). 
	    As we discussed in Section~\ref{sec:iso} this could be due to the diffusion not been taken into account close to the edge of the wind in the analytical model.
	    Later on, as in the non-evolving wind case the luminosity continues to decay with time but now following a shallower power law $t^{-1/6}$ caused by the growing importance of diffusion. 
	    Notice that the model with the shortest characteristic timescale $v_{\rm w}t'/r_{\rm in}=10^{-2}$ (connected red dots) does not manage to get into this phase, instead it skips directly into the next one. 
	    At $t>t'$ the wind quickly evolves towards becoming optically thin. 
	    This is shown as a rise following $\propto t^{10/9}$ due to our choice of $\beta=5/3$. 
	    Once the wind becomes completely optically thin, the luminosity is no longer attenuated so it increases up to its maximum level $L_{\rm obs}/L_*\approx 1$ and remains constant. 
	    At the beginning of this stage, we observe some wiggles and slight overestimation of the reprocessed luminosity, which are due to the small fluctuations appearing in transitions between optically-thick to optically-thin media. 
	    Overall, the evolution agrees very well with the analytical scaling relations.
	
	\subsection{Anisotropic irradiation with constant wind}
	
		We now consider a wind with constant $A$ but with an anisotropic irradiation source characterised by an opening angle along the polar coordinate $\theta_{\rm op}$, which is measured from a polar axis. 
		As a result, the radiation is injected in a cone. 
		In a real astrophysical object, this geometry could arise, for example, when a thick disk blocks radiation from propagating near the equatorial plane.
		In Figure~\ref{fig:maps}, we present radiation energy density maps at four different times for a model with $A=10^2$ and $\theta_{\rm op}=15\degr$. 
		Notice that the spatial as well as the radiation energy density scales change with time due to the grid radial motion.  
		In the top-left panel, it is possible to observe how radiation is injected solely at small polar angles. 
		The top-right panel shows how radiation starts to diffuse within the wind to larger polar angles. 
		However, at such short timescales radiation has not had enough time to diffuse far.
		As the system evolves, the radiation map looks more evenly distributed as we see in the bottom-left panel. 
		However, at small scales it is still possible to observe the signatures of radiation injection only within the polar cone. 
		Energy density in the bottom-right panel looks like a completely isotropic source, because the wind became more transparent in its outer regions and the radiation diffused easily in the polar direction.
		
		Figure~\ref{fig:asym} shows the simulated light curves for models with $A=10^2$: $\theta_{\rm op}=15\degr$ (left panel) and $\theta_{\rm op}=45\degr$ (right panel), along four different lines of sight: $\theta_\text{obs}=0\degr$ (blue dots), $30\degr$ (orange dots), $60\degr$ (green dots), and $90\degr$ (red dots). 
		In these cases, $L_{\rm obs}$ was calculated as the isotropic equivalent bolometric luminosity. 
		To obtain it we had to remap our simulations into three-dimensional space by assuming azimuthal symmetry and reflective symmetry in the polar direction.
		Then, we integrated the total radiated power at the photosphere using the line-of-sight projection of the radiative flux. 
		In Figure~\ref{fig:asym}, we also included the simulated light curve of with isotropic radiation and $A=10^2$ (solid black line) as a reference.
		At early times ($v_\text{w}t/r_\text{in}\lesssim1$), the light curves of both models display behave analogously but scaled according to the projection effect of the rays from the illuminated region into the different lines of sight. 
		Notice that if the line of sight is along or close to the irradiated area an observer would tend to overestimate the luminosity due to the assumption of an isotropic source. 
		The smaller the confinement angle $\theta_{\rm op}$ the larger the overestimation. 
	    In addition, the case with a smaller $\theta_{\rm op}$ shows a wider spread of scaling between the light curves compared to the case with larger $\theta_{\rm op}$. 
	    This is simply as the projection effect is starker if the radiation is confined into a smaller angle cone when comparing the light curves along the most extreme lines of sight.
	    Further in the evolution, light curves in both models have a shallower power-law decay similar to the isotropic irradiation case. 
	    During this phase, the radiation from the directly illuminated region starts to diffuse to larger polar angles. 
		As a result, these light curves start to differ from the isotropic irradiation case. 
		In the case with a smaller $\theta_{\rm op}$ the differences are much more noticeable, especially at high inclination lines of sight. 
		In the other case, although differences are mild it is also possible to observe a slightly increasing trend after reaching the minimum value in the light curves at high inclination.  
		Although behaviours in both cases are different, the underlying cause is the same. 
		Radiation needs enough time for being able to reach to higher polar angles, and of course the higher the line-of-sight the longer the longer it takes for radiation to travel. 
        
        In order to quantify the timescale for light curves along different lines of sight for reaching this phase we estimated the diffuse timescale in the polar angle direction,
		\begin{equation}
		    t_{\text{dif},\theta}\sim\frac{(r\Delta\theta)\tau}{c}=\frac{\rho k_{\rm s} (r\Delta\theta)^2}{c}=\frac{\dot{M}k_{\rm s}\Delta\theta^2}{4\pi cv_{\rm w}},
		\end{equation}
		\noindent where $\Delta\theta = |\theta_\text{op} - \theta_\text{obs}|$ corresponds to the polar angle difference between a given line of sight and the irradiation opening angle $\theta_{\rm op}$. 
		This quantity is shown as vertical dashed lines in Figure~\ref{fig:asym}, where the colours represent also the line-of-sight angle use to compute $t_{\text{dif},\theta}$. 
		Approximately, after such timescales radiation has had enough time to diffuse to higher inclinations so that light curves resemble qualitatively the isotropic irradiation case. 
		Finally, we observe that at late times all light curves converge as at $t\gg t_{\text{dif},\theta}$ the source must behave as an isotropic source coinciding with the isotropic irradiation model. 
		
		In summary, the anisotropic irradiation of the wind can change significantly the simple sequence of analytic power-law scalings if the system is observed along extreme lines of sight compared to the irradiation cone angular size. 
		We expect additional complications when considering that both the irradiation and the wind are anisotropic, which is likely to occur in tidal disruption events and similar extreme situations.
	   
\section{Conclusions}
\label{sec:conclusions}

    We have developed a new module for radiation treatment and coupling with hydrodynamics for the moving-mesh hydrodynamic code JET. 
    Our code solves the equations of radiative hydrodynamics in the mixed-frame formulation using the FLD approximation as a closure relation. 
    We verified its capability of performing multi-dimensional simulations over many orders of magnitude in both space and time, which makes it an ideal tool for modelling astrophysical transients. 
    As a first application, we have performed two-dimensional moving-mesh radiation hydrodynamic simulations of an optically-thick wind reprocessing irradiation from a central source. 
    When both the wind and the irradiating source are spherically-symmetric, the simulated light curves are overall in agreement with the scaling relations derived analytically by \cite{P20} for both the constant and time-dependent winds.  
    We found differences at the earliest stage of the evolution as neglecting completely the diffusion close to the edge of the wind might not be completely justified. 

    As a new result made possible by the multi-dimensional capability of our code, we also studied the case of anisotropic irradiation with a constant wind. 
    Here, our results suggest that the evolution is qualitatively similar to the isotropic source as long as we observe either from the direction of the radiation injection or if the opening angle of the irradiation is large enough ($\theta_{\rm op}\gtrsim 45\degr$). 
    Differences become important when we observe outside of this region as radiation needs enough time to diffuse through the wind to larger polar angles before it can escape from the wind. 
    Depending on the opening angle of the irradiation, the differences in the inferred luminosity could be up to one order of magnitude. 
    Although definitive results require more sophisticated method of calculating the light curves, our results suggest that diffusion in the polar direction can modify the sequence of power laws seen in the fiducial isotropic irradiation scenario.
    
    In this work, we have restricted our analysis to an ideal scenario in order to make a direct comparison with analytical predictions. 
    Nevertheless, our numerical tool is certainly capable of dealing with much more complex and realistic problems. 
    A next step would be the use of more realistic opacity prescriptions so that it is possible to reproduce observational data. 
    In the case of tidal disruption events, we plan to investigate models with much more complex geometries for the entire duration of the observed transient so that we can synthesise their light curves \citep{G15,metzger16,bonnerot21}. 
    In the case of type-IIn supernovae, despite recent work \citep[e.g.][]{S16,S19} further modelling is needed for understanding observational signatures of different geometries of CSM, e.g. slabs resulting from the wind-wind collisions of their precursors \citep{kurfurst20}. 
    Also, it is necessary to simulate observational signatures of the final stage of stellar mergers in order to test their relation with light curves of luminous red novae. 
    Additional physical processes might need to be considered in order to capture the full picture of such phenomena. 
    The further addition of dust formation \citep[e.g.][]{Z14} or even chemical networks may be crucial for modelling properly the evolution of some transients with low inferred temperatures ($<1000\rm\ K$). 
    The Lagrangian nature of our code makes it ideal for implementing similar physical effects.

\section*{Acknowledgements}
    We would like to thank the anonymous referee for useful comments and suggestions that improved this article. 
    We also thank Jaroslav Hron for constructive discussions about iterative linear solvers and preconditioners. 
    Furthermore, we thank Cl\'ement Bonnerot for his comments on the early version of draft. 
    The research of DC and OP has been supported by Horizon 2020 ERC Starting Grant `Cat-In-hAT' (grant agreement no. 803158). 
    This work made use of \textsc{python} libraries \textsc{numpy} \citep{harris20} and \textsc{matplotlib} \citep{hunter07}, as well as of the NASA’s Astrophysics Data System.

\section*{Data Availability}

     The output files from our simulations will be shared on reasonable request to the corresponding author.
     
%%%%%%%%%%%%%%%%%%%%%%%%%%%%%%%%%%%%%%%%%%%%%%%%%%



%\bibliographystyle{mnras}

\begin{thebibliography}{99}
%    \scriptsize
	\bibitem[\protect\citeauthoryear{Alme \& Wilson}{1973}]{A73} Alme M.~L., Wilson J.~R., 1973, \apj, 186, 1015. doi:10.1086/152566
	\bibitem[\protect\citeauthoryear{Andrews \& Smith}{2018}]{andrews18} Andrews J.~E., Smith N., 2018, \mnras, 477, 74. doi:10.1093/mnras/sty584
	\bibitem[\protect\citeauthoryear{Aydi et al.}{2020}]{aydi20} Aydi E., Sokolovsky K.~V., Chomiuk L., Steinberg E., Li K.~L., Vurm I., Metzger B.~D., et al., 2020, NatAs, 4, 776. doi:10.1038/s41550-020-1070-y
	\bibitem[\protect\citeauthoryear{Balay et al.}{2020}]{B20} Balay, S., et al., 2020, {PETS}c {W}eb page, https://www.mcs.anl.gov/petsc
	\bibitem[\protect\citeauthoryear{Bersten, Benvenuto, \& Hamuy}{2011}]{bersten11} Bersten M.~C., Benvenuto O., Hamuy M., 2011, \apj, 729, 61. doi:10.1088/0004-637X/729/1/61
	\bibitem[\protect\citeauthoryear{Bilinski et al.}{2018}]{bilinski18} Bilinski C., Smith N., Williams G.~G., Smith P., Zheng W., Graham M.~L., Mauerhan J.~C., et al., 2018, MNRAS, 475, 1104. doi:10.1093/mnras/stx3214
	\bibitem[\protect\citeauthoryear{Blagorodnova et al.}{2021}]{blagorodnova21} Blagorodnova N., Klencki J., Pejcha O., Vreeswijk P.~M., Bond H.~E., Burdge K.~B., De K., et al., 2021, arXiv, arXiv:2102.05662
	\bibitem[\protect\citeauthoryear{Blinnikov et al.}{1998}]{blinnikov98} Blinnikov S.~I., Eastman R., Bartunov O.~S., Popolitov V.~A., Woosley S.~E., 1998, \apj, 496, 454. doi:10.1086/305375
	\bibitem[\protect\citeauthoryear{Blondin, Lundqvist, \& Chevalier}{1996}]{blondin96} Blondin J.~M., Lundqvist P., Chevalier R.~A., 1996, \apj, 472, 257. doi:10.1086/178060
	\bibitem[\protect\citeauthoryear{Bonnerot, Lu, \& Hopkins}{2021}]{bonnerot21} Bonnerot C., Lu W., Hopkins P.~F., 2021, \mnras, 504, 4885. doi:10.1093/mnras/stab398
	\bibitem[\protect\citeauthoryear{Chatzopoulos \& Weide}{2019}]{C19} Chatzopoulos E., Weide K., 2019, \apj, 876, 148. doi:10.3847/1538-4357/ab18f9
	\bibitem[\protect\citeauthoryear{Chomiuk et al.}{2014}]{chomiuk14} Chomiuk L., Linford J.~D., Yang J., O'Brien T.~J., Paragi Z., Mioduszewski A.~J., Beswick R.~J., et al., 2014, \nat, 514, 339. doi:10.1038/nature13773
	\bibitem[\protect\citeauthoryear{Chomiuk, Metzger, \& Shen}{2020}]{chomiuk2020} Chomiuk L., Metzger B.~D., Shen K.~J., 2020, arXiv, arXiv:2011.08751
	\bibitem[\protect\citeauthoryear{Chugai \& Danziger}{1994}]{chugai94} Chugai N.~N., Danziger I.~J., 1994, \mnras, 268, 173. doi:10.1093/mnras/268.1.173
    \bibitem[\protect\citeauthoryear{Commer{\c{c}}on et al.}{2011}]{C11} Commer{\c{c}}on B., Teyssier R., Audit E., Hennebelle P., Chabrier G., 2011, \aap, 529, A35. doi:10.1051/0004-6361/201015880
	\bibitem[\protect\citeauthoryear{Dai, Lodato, \& Cheng}{2021}]{D21} Dai J.~L., Lodato G., Cheng R., 2021, SSRv, 217, 12. doi:10.1007/s11214-020-00747-x
	\bibitem[\protect\citeauthoryear{Dessart \& Hillier}{2005}]{dessart05} Dessart L., Hillier D.~J., 2005, \aap, 437, 667. doi:10.1051/0004-6361:20042525
	\bibitem[\protect\citeauthoryear{Dessart \& Audit}{2019}]{dessart19} Dessart L., Audit E., 2019, \aap, 629, A17. doi:10.1051/0004-6361/201935794
	\bibitem[\protect\citeauthoryear{Duffell \& MacFadyen}{2011}]{D11} Duffell P.~C., MacFadyen A.~I., 2011, \apjs, 197, 15. doi:10.1088/0067-0049/197/2/15
	\bibitem[\protect\citeauthoryear{Duffell \& MacFadyen}{2013}]{D13} Duffell P.~C., MacFadyen A.~I., 2013, \apj, 775, 87. doi:10.1088/0004-637X/775/2/87
	\bibitem[\protect\citeauthoryear{Falgout et al.}{2006}]{F06} Falgout R.~D., Jones J.~E, Yang U.~M., 2006, in Bruaset, A.~M., Tveito, A., ed., Numerical Solution of Partial Differential Equations on Parallel Computers, Lecture Notes in Computational Science and Engineering, vol 51., Springer-Verlag, Berlin, p. 267
	\bibitem[\protect\citeauthoryear{Fang et al.}{2020}]{fang20} Fang K., Metzger B.~D., Vurm I., Aydi E., Chomiuk L., 2020, \apj, 904, 4. doi:10.3847/1538-4357/abbc6e
	\bibitem[\protect\citeauthoryear{Gittings et al.}{2008}]{gittings08} Gittings M., Weaver R., Clover M., Betlach T., Byrne N., Coker R., Dendy E., et al., 2008, CS\&D, 1, 015005. doi:10.1088/1749-4699/1/1/015005
	\bibitem[\protect\citeauthoryear{Gonz{\'a}lez, Audit, \& Huynh}{2007}]{gonzalez07} Gonz{\'a}lez M., Audit E., Huynh P., 2007, \aap, 464, 429. doi:10.1051/0004-6361:20065486
	\bibitem[\protect\citeauthoryear{Guillochon \& Ramirez-Ruiz}{2015}]{G15} Guillochon J., Ramirez-Ruiz E., 2015, \apj, 809, 166. doi:10.1088/0004-637X/809/2/166
	\bibitem[\protect\citeauthoryear{Harris et al.}{2020}]{harris20} Harris C.~R., et al., 2020, Nature, 585, 357. doi:10.1038/s41586-020-2649-2
	\bibitem[\protect\citeauthoryear{Hunter et al.}{2007}]{hunter07} Hunter J.~D., 2007, Computing in Science \& Engineering, 9, 90. 
	doi: 10.1109/MCSE.2007.55
	\bibitem[\protect\citeauthoryear{Ivanova et al.}{2013}]{ivanova13} Ivanova N., Justham S., Avendano Nandez J.~L., Lombardi J.~C., 2013, \sci, 339, 433. doi:10.1126/science.1225540
	\bibitem[\protect\citeauthoryear{Kasen, Thomas, \& Nugent}{2006}]{kasen06} Kasen D., Thomas R.~C., Nugent P., 2006, \apj, 651, 366. doi:10.1086/506190
	\bibitem[\protect\citeauthoryear{Kasen et al.}{2017}]{kasen17} Kasen D., Metzger B., Barnes J., Quataert E., Ramirez-Ruiz E., 2017, Natur, 551, 80. doi:10.1038/nature24453
	\bibitem[\protect\citeauthoryear{Kasen \& Woosley}{2009}]{kasen09} Kasen D., Woosley S.~E., 2009, \apj, 703, 2205. doi:10.1088/0004-637X/703/2/2205
	\bibitem[\protect\citeauthoryear{Kerzendorf \& Sim}{2014}]{kerzendorf14} Kerzendorf W.~E., Sim S.~A., 2014, \mnras, 440, 387. doi:10.1093/mnras/stu055
	\bibitem[\protect\citeauthoryear{Kolb et al.}{2013}]{K13} Kolb S.~M., Stute M., Kley W., Mignone A., 2013, \aap, 559, A80. doi:10.1051/0004-6361/201321499
	\bibitem[\protect\citeauthoryear{Krumholz et al.}{2007}]{K07} Krumholz M.~R., Klein R.~I., McKee C.~F., Bolstad J., 2007, \apj, 667, 626. doi:10.1086/520791
	\bibitem[\protect\citeauthoryear{Kurf{\"u}rst \& Krti{\v{c}}ka}{2019}]{kurfurst19} Kurf{\"u}rst P., Krti{\v{c}}ka J., 2019, \aap, 625, A24. doi:10.1051/0004-6361/201833429
	\bibitem[\protect\citeauthoryear{Kurf{\"u}rst, Pejcha, \& Krti{\v{c}}ka}{2020}]{kurfurst20} Kurf{\"u}rst P., Pejcha O., Krti{\v{c}}ka J., 2020, \aap, 642, A214. doi:10.1051/0004-6361/202039073
	\bibitem[\protect\citeauthoryear{Leonard et al.}{2000}]{leonard00} Leonard D.~C., Filippenko A.~V., Barth A.~J., Matheson T., 2000, \apj, 536, 239. doi:10.1086/308910
	\bibitem[\protect\citeauthoryear{Levermore \& Pomraning}{1981}]{L81} Levermore C.~D., Pomraning G.~C., 1981, \apj, 248, 321. doi:10.1086/159157
	\bibitem[\protect\citeauthoryear{Li et al.}{2017}]{li17} Li K.-L., Metzger B.~D., Chomiuk L., Vurm I., Strader J., Finzell T., Beloborodov A.~M., et al., 2017, NatAs, 1, 697. doi:10.1038/s41550-017-0222-1
	\bibitem[\protect\citeauthoryear{Margutti et al.}{2019}]{margutti19} Margutti R., Metzger B.~D., Chornock R., Vurm I., Roth N., Grefenstette B.~W., Savchenko V., et al., 2019, \apj, 872, 18. doi:10.3847/1538-4357/aafa01
	\bibitem[\protect\citeauthoryear{Mauerhan et al.}{2013}]{mauerhan13} Mauerhan J.~C., Smith N., Silverman J.~M., Filippenko A.~V., Morgan A.~N., Cenko S.~B., Ganeshalingam M., et al., 2013, \mnras, 431, 2599. doi:10.1093/mnras/stt360
	\bibitem[\protect\citeauthoryear{McDowell, Duffell, \& Kasen}{2018}]{mcdowell18} McDowell A.~T., Duffell P.~C., Kasen D., 2018, \apj, 856, 29. doi:10.3847/1538-4357/aaa96e
	\bibitem[\protect\citeauthoryear{Metzger et al.}{2014}]{metzger14} Metzger B.~D., Hasco{\"e}t R., Vurm I., Beloborodov A.~M., Chomiuk L., Sokoloski J.~L., Nelson T., 2014, \mnras, 442, 713. doi:10.1093/mnras/stu844
	\bibitem[\protect\citeauthoryear{Metzger \& Pejcha}{2017}]{metzger17} Metzger B.~D., Pejcha O., 2017, \mnras, 471, 3200. doi:10.1093/mnras/stx1768
	\bibitem[\protect\citeauthoryear{Metzger \& Stone}{2016}]{metzger16} Metzger B.~D., Stone N.~C., 2016, \mnras, 461, 948. doi:10.1093/mnras/stw1394
	\bibitem[\protect\citeauthoryear{Mignone et al.}{2007}]{M07} Mignone A., Bodo G., Massaglia S., Matsakos T., Tesileanu O., Zanni C., Ferrari A., 2007, \apjs, 170, 228. doi:10.1086/513316
	\bibitem[\protect\citeauthoryear{Mihalas \& Mihalas}{1999}]{mihalas1999} Mihalas D., Mihalas, B. W., 1999, Foundations of Radiation Hydrodynamics (New York: Dover)
	\bibitem[\protect\citeauthoryear{Moens et al.}{2021}]{moens21} Moens M., Sundqvist J.~O., El Mellah I., Poniatowski L., Teunissen J., Keppens R., 2021, arXiv, arXiv:2104.03968 
	\bibitem[\protect\citeauthoryear{Morozova et al.}{2015}]{morozova15} Morozova V., Piro A.~L., Renzo M., Ott C.~D., Clausen D., Couch S.~M., Ellis J., et al., 2015, \apj, 814, 63. doi:10.1088/0004-637X/814/1/63
	\bibitem[\protect\citeauthoryear{Paxton et al.}{2011}]{paxton11} Paxton B., Bildsten L., Dotter A., Herwig F., Lesaffre P., Timmes F., 2011, \apjs, 192, 3. doi:10.1088/0067-0049/192/1/3
    \bibitem[\protect\citeauthoryear{Paxton et al.}{2018}]{paxton18} Paxton B., Schwab J., Bauer E.~B., Bildsten L., Blinnikov S., Duffell P., Farmer R., et al., 2018, \apjs, 234, 34. doi:10.3847/1538-4365/aaa5a8
	\bibitem[\protect\citeauthoryear{Paxton et al.}{2019}]{paxton19} Paxton B., Smolec R., Schwab J., Gautschy A., Bildsten L., Cantiello M., Dotter A., et al., 2019, \apjs, 243, 10. doi:10.3847/1538-4365/ab2241
	\bibitem[\protect\citeauthoryear{Pejcha}{2014}]{pejcha14} Pejcha O., 2014, \apj, 788, 22. doi:10.1088/0004-637X/788/1/22
	\bibitem[\protect\citeauthoryear{Pejcha et al.}{2017}]{pejcha17} Pejcha O., Metzger B.~D., Tyles J.~G., Tomida K., 2017, \apj, 850, 59. doi:10.3847/1538-4357/aa95b9
	\bibitem[\protect\citeauthoryear{Piro \& Lu}{2020}]{P20} Piro A.~L., Lu W., 2020, \apj, 894, 2. doi:10.3847/1538-4357/ab83f6
	\bibitem[\protect\citeauthoryear{Pumo \& Zampieri}{2011}]{pumo11} Pumo M.~L., Zampieri L., 2011, \apj, 741, 41. doi:10.1088/0004-637X/741/1/41
	\bibitem[\protect\citeauthoryear{Ramsey, Clarke, \& Men'shchikov}{2012}]{R12} Ramsey J.~P., Clarke D.~A., Men'shchikov A.~B., 2012, \apjs, 199, 13. doi:10.1088/0067-0049/199/1/13
	\bibitem[\protect\citeauthoryear{Ramsey \& Dullemond}{2015}]{R15} Ramsey J.~P., Dullemond C.~P., 2015, \aap, 574, A81. doi:10.1051/0004-6361/201424954
	\bibitem[\protect\citeauthoryear{Rivera-Paleo \& Guzm{\'a}n}{2019}]{rivera-paleo19} Rivera-Paleo F.~J., Guzm{\'a}n F.~S., 2019, ApJS, 241, 28. doi:10.3847/1538-4365/ab0d8c
	\bibitem[\protect\citeauthoryear{Roth \& Kasen}{2015}]{roth15} Roth N., Kasen D., 2015, \apjs, 217, 9. doi:10.1088/0067-0049/217/1/9
	\bibitem[\protect\citeauthoryear{Shen, Nakar, \& Piran}{2016}]{shen16} Shen R.-F., Nakar E., Piran T., 2016, \mnras, 459, 171. doi:10.1093/mnras/stw645
	\bibitem[\protect\citeauthoryear{Smith}{2013}]{smith13} Smith N., 2013, \mnras, 434, 102. doi:10.1093/mnras/stt1004
	\bibitem[\protect\citeauthoryear{Smith et al.}{2015}]{smith15} Smith N., Mauerhan J.~C., Cenko S.~B., Kasliwal M.~M., Silverman J.~M., Filippenko A.~V., Gal-Yam A., et al., 2015, \mnras, 449, 1876. doi:10.1093/mnras/stv354
	\bibitem[\protect\citeauthoryear{Soker \& Tylenda}{2003}]{soker03} Soker N., Tylenda R., 2003, ApJL, 582, L105. doi:10.1086/367759
	\bibitem[\protect\citeauthoryear{Sukhbold et al.}{2016}]{sukhbold16} Sukhbold T., Ertl T., Woosley S.~E., Brown J.~M., Janka H.-T., 2016, \apj, 821, 38. doi:10.3847/0004-637X/821/1/38
	\bibitem[\protect\citeauthoryear{Suzuki, Maeda, \& Shigeyama}{2016}]{S16} Suzuki A., Maeda K., Shigeyama T., 2016, \apj, 825, 92. doi:10.3847/0004-637X/825/2/92
	\bibitem[\protect\citeauthoryear{Suzuki, Moriya, \& Takiwaki}{2019}]{S19} Suzuki A., Moriya T.~J., Takiwaki T., 2019, \apj, 887, 249. doi:10.3847/1538-4357/ab5a83
	\bibitem[\protect\citeauthoryear{Teyssier}{2002}]{T02} Teyssier R., 2002, \aap, 385, 337. doi:10.1051/0004-6361:20011817
	\bibitem[\protect\citeauthoryear{Turner \& Stone}{2001}]{T01} Turner N.~J., Stone J.~M., 2001, \apjs, 135, 95. doi:10.1086/321779
	\bibitem[\protect\citeauthoryear{Tylenda \& Soker}{2006}]{tylenda06} Tylenda R., Soker N., 2006, \aap, 451, 223. doi:10.1051/0004-6361:20054201
	\bibitem[\protect\citeauthoryear{Uno \& Maeda}{2020a}]{uno20} Uno K., Maeda K., 2020a, \apj, 897, 156. doi:10.3847/1538-4357/ab9632
	\bibitem[\protect\citeauthoryear{Uno \& Maeda}{2020b}]{uno20b} Uno K., Maeda K., 2020b, ApJL, 905, L5. doi:10.3847/2041-8213/abca32
	\bibitem[\protect\citeauthoryear{van der Vorst}{1992}]{V92} van der Vorst H.~A., 1992, SIAM Journal on Scientific Computing, 13, 631
	\bibitem[\protect\citeauthoryear{van Marle et al.}{2010}]{vanmarle10} van Marle A.~J., Smith N., Owocki S.~P., van Veelen B., 2010, \mnras, 407, 2305. doi:10.1111/j.1365-2966.2010.16851.x
	\bibitem[\protect\citeauthoryear{Villar et al.}{2017}]{villar17} Villar V.~A., Guillochon J., Berger E., Metzger B.~D., Cowperthwaite P.~S., Nicholl M., Alexander K.~D., et al., 2017, ApJL, 851, L21. doi:10.3847/2041-8213/aa9c84
	\bibitem[\protect\citeauthoryear{Vlasis, Dessart, \& Audit}{2016}]{vlasis16} Vlasis A., Dessart L., Audit E., 2016, \mnras, 458, 1253. doi:10.1093/mnras/stw410
	\bibitem[\protect\citeauthoryear{Woosley \& Heger}{2007}]{woosley07} Woosley S.~E., Heger A., 2007, PhR, 442, 269. doi:10.1016/j.physrep.2007.02.009
	\bibitem[\protect\citeauthoryear{Yang \& Henson}{2002}]{Y02} Yang U.~M., Henson V.~E., 2002, Applied Numerical Mathematics, 41, 155,
	\bibitem[\protect\citeauthoryear{Zhang et al.}{2011}]{Z11} Zhang W., Howell L., Almgren A., Burrows A., Bell J., 2011, \apjs, 196, 20. doi:10.1088/0067-0049/196/2/20
	\bibitem[\protect\citeauthoryear{Zhang et al.}{2013}]{zhang13} Zhang W., Howell L., Almgren A., Burrows A., Dolence J., Bell J., 2013, \apjs, 204, 7. doi:10.1088/0067-0049/204/1/7
	\bibitem[\protect\citeauthoryear{Zhu et al.}{2014}]{Z14} Zhu Z., Stone J.~M., Rafikov R.~R., Bai X.-. ning ., 2014, \apj, 785, 122. doi:10.1088/0004-637X/785/2/122
\end{thebibliography}


%%%%%%%%%%%%%%%%%%%%%%%%%%%%%%%%%%%%%%%%%%%%%%%%%

%%%%%%%%%%%%%%%%% APPENDICES %%%%%%%%%%%%%%%%%%%%%
\onecolumn
\appendix

\section{Flux-limited diffusion in JET}
\label{sec:app1}

	Our new module for radiation treatment in JET considers the radiation hydrodynamics equations in the mixed-frame formulation so that we can ensure the energy conservation to high precision \citep{K07}. 
	We work with frequency-integrated quantities, and assume local thermodynamic equilibrium. 
	Then, the equations keeping terms up to order $\mathcal{O}(u/c)$ \citep{Z11} are given by
	\begin{eqnarray}
		\frac{\partial \rho}{\partial t}+\nabla\cdot\left(\rho\bf{u}\right) & = & 0,\\
		\frac{\partial (\rho\mathbf{u})}{\partial t}+\nabla\cdot\left(\rho\bf{u}\bf{u}\right) + \nabla p & = & \frac{\chi_{\rm F}}{c}\mathbf{F}_\text{r}^{(0)} -\kappa_{\rm P}\left(\frac{\mathbf{u}}{c}\right)\left(a_{\rm r}T^4-E_\text{r}^{(0)}\right),\\
		\frac{\partial(\rho E)}{\partial t}+\nabla\cdot\left(\rho E\mathbf{u}+p\mathbf{u}\right) & = & \chi_{\rm F}\left(\frac{\mathbf{u}}{c}\right)\cdot\mathbf{F}_\text{r}^{(0)}\nonumber-c\kappa_{\rm P}\left(a_{\rm r}T^4-E_\text{r}^{(0)}\right),\\
		\frac{\partial E_\text{r}}{\partial t}+\nabla\cdot\mathbf{F}_\text{r} & = & -\chi_{\rm F}\left(\frac{\mathbf{u}}{c}\right)\cdot\mathbf{F}_\text{r}^{(0)}\nonumber+c\kappa_{\rm P}\left(a_{\rm r}T^4-E_\text{r}^{(0)}\right),\\
		\frac{1}{c^2}\frac{\partial\mathbf{F}_\text{r}}{\partial t}+\nabla\cdot{\rm P}_\text{r} & = & -\frac{\chi_{\rm F}}{c}\mathbf{F}_\text{r}^{(0)}+\kappa_{\rm P}\left(\frac{\mathbf{u}}{c}\right)\left(a_{\rm r}T^4-E_\text{r}^{(0)}\right),
	\end{eqnarray}
	
	\noindent where $(\rho,\mathbf{u},p)$ correspond to the primitive hydrodynamic variables: mass density, fluid velocity, and thermal pressure, respectively. 
	In addition, $E$ and $T$ are the total energy of matter per unit mass (internal plus kinetic) and temperature of the fluid.
	The radiation variables $(E_\text{r},\mathbf{F}_\text{r},{\rm P}_\text{r})$ are the radiation energy density, radiation flux, and radiation pressure tensor, respectively. 
	The constants $c$ and $a_\text{r}$ represent the speed of light and the radiation constant. 
	The quantities $\kappa_{\rm P}$ and $\chi_{\rm F}$ are the Planck mean and flux mean interaction coefficients, which have units of inverse length. 
	They are related to the Planck mean opacity $k_\text{P}$ and flux mean opacity $k_\text{F}$ through $\kappa_\text{P}=\rho k_\text{P}$ and $\chi_\text{F}=\rho k_\text{F}$, respectively. 
	Thus, opacities have units of square length divided by mass (e.g. $\rm cm^2\ g^{-1}$).
	Variables with the superscript $(0)$ are measured in the co-moving frame while variables without it are measured in the lab frame.
	
	Notice that we have six variables but a set of only five equations. 
	In order to solve the system it is necessary to make an extra assumption in order to close the system. 
	In this case, we consider the flux-limited diffusion (herefter FLD) approximation \citep{A73} so that the co-moving radiation flux can be written as a function of the co-moving radiation energy density following the Fick's law as follows
	\begin{eqnarray}
		\mathbf{F}_\text{r}^{(0)}=-\frac{c\lambda}{\chi_{\rm R}}\nabla E_\text{r}^{(0)},
	\end{eqnarray}
	\noindent where $\chi_{\rm R}$ is the Rosseland mean of the sum of absorption and scattering coefficients. 
	This value is related to the Rosseland mean opacity $k_\text{R}$ through $\chi_R=\rho k_\text{R}$.
	Also, $\lambda$ represents the flux limiter, which is given by \cite{L81},
	\begin{eqnarray}
		\lambda=\frac{2+R}{6+3R+R^2},
	\end{eqnarray}
	\noindent where
	\begin{eqnarray}
		R=\frac{|\nabla E_\text{r}^{(0)}|}{\chi_{\rm R}E_\text{r}^{(0)}}.
	\end{eqnarray}
	\noindent
	The flux limiter allows radiation to diffuse in the optically thick regime so $\lambda\to 1/3$, while in the optically-thin limit $\lambda\to 1/R$ so that radiation propagates with the speed of light. 
	Within the FLD approximation the radiation pressure is ${\rm P}_\text{r}^{(0)}=f^{(0)}E_\text{r}^{(0)}$, where $f$ is the Eddington factor, and is given by $f=\lambda+\lambda^2R^2$.
	
	We make the assumption that $\chi_{\rm F}\equiv \chi_{\rm R}$, which is valid in the optically thick regime \citep{mihalas1999}. 
	Furthermore, we consider that the flux limiter depends on radiation energy density measured in the lab frame \citep{C19}. 
	Then, we arrive to the radiation hydrodynamics equations in the FLD approximation up to order $\mathcal{O}(u/c)$
	\begin{eqnarray}
		\frac{\partial \rho}{\partial t} + \nabla\cdot\left(\rho\mathbf{u}\right) & = & 0,\\
		\frac{\partial (\rho\mathbf{u})}{\partial t}+\nabla\cdot\left(\rho\mathbf{u}\mathbf{u}\right) + \nabla p + \lambda\nabla E_\text{r} & = & \bm{0},\\
		\frac{\partial(\rho E)}{\partial t}+\nabla\cdot\left(\rho E\mathbf{u}+p\mathbf{u}\right) + \lambda\mathbf{u}\cdot\nabla E_\text{r}& = & -c\kappa_{\rm P}\left(a_{\rm r}T^4-E_\text{r}^{(0)}\right),\\
		\frac{\partial E_\text{r}}{\partial t}+\nabla\cdot\left(\frac{3-f}{2}E_\text{r}\mathbf{u}\right)-\lambda\mathbf{u}\cdot\nabla E_\text{r} & = & c\kappa_{\rm P}\left(a_{\rm r}T^4-E_\text{r}^{(0)}\right)+\nabla\cdot\left(\frac{c\lambda}{\chi_{\rm R}}\nabla E_\text{r}\right),
	\end{eqnarray} 
	\noindent
	where we have dropped terms that are not significant at leading order in the streaming, static diffusion, or dynamic diffusion regimes \citep[see Section~2.2 in][for a discussion]{K07}. 
	The co-moving and lab frame radiation quantities are related through the expressions derived by \cite{Z11},
	\begin{eqnarray}
		E_\text{r}^{(0)}=E_\text{r}-\frac{2}{c^2}\mathbf{u}\cdot\mathbf{F}_\text{r}^{(0)}+\mathcal{O}(u^2/c^2)=E_\text{r}+2\frac{\lambda}{\chi_{\rm R}}\frac{\mathbf{u}}{c}\cdot\nabla E_\text{r}+\mathcal{O}(u^2/c^2).
		\label{eq:frame}
	\end{eqnarray}
	
	In order to solve the system of equations we start focusing on the hyperbolic subsystem of equations,
	\begin{eqnarray}
		\frac{\partial \rho}{\partial t} + \nabla\cdot\left(\rho\mathbf{u}\right) & = & 0,\\
		\frac{\partial (\rho\mathbf{u})}{\partial t}+\nabla\cdot\left(\rho\mathbf{u}\mathbf{u}\right) + \nabla p + \lambda\nabla E_\text{r} & = & \bm{0},\\
		\frac{\partial(\rho E)}{\partial t}+\nabla\cdot\left(\rho E\mathbf{u}+p\mathbf{u}\right) + \lambda\mathbf{u}\cdot\nabla E_\text{r}& = & 0,\\
		\frac{\partial E_\text{r}}{\partial t}+\nabla\cdot\left(\frac{3-f}{2}E_\text{r}\mathbf{u}\right)-\lambda\mathbf{u}\cdot\nabla E_\text{r} & = & 0.
	\end{eqnarray} 
	Following \cite{Z11} we solve this system making use of the Godunov method through a characteristic-based Riemann solver. 
	The eigenvalues of the problem are $u-c_s,u,u+c_s$, where $c_s$ is the radiation-modified sound speed given by
	\begin{eqnarray}
		c_s=\sqrt{\gamma\frac{p}{\rho}+(\lambda+1)\frac{\lambda E_\text{r}}{\rho}}.
	\end{eqnarray}
	Here, $\gamma$ is the adiabatic index and we have assumed that $(3-f)/2=\lambda+1$. 
	Rearranging the system of equations we obtain an equation for the total energy (matter plus radiation) conservation, i.e.
	\begin{eqnarray}
		\frac{\partial \rho}{\partial t} + \nabla\cdot\left(\rho\mathbf{u}\right) & = & 0,\\
		\frac{\partial (\rho\mathbf{u})}{\partial t}+\nabla\cdot\left(\rho\bf{u}\bf{u}\right) + \nabla p & = & \bm{0},\\
		\frac{\partial(\rho E+E_\text{r})}{\partial t}+\nabla\cdot\left(\rho E\mathbf{u}+p\mathbf{u}+\frac{3-f}{2}E_\text{r}\mathbf{u}\right) & = & 0,\\
		\frac{\partial E_\text{r}}{\partial t}+\nabla\cdot\left(\frac{3-f}{2}E_\text{r}\mathbf{u}\right)& = & 0.
	\end{eqnarray}
	
	Once we solve the hyperbolic portion of the system of equations we need to deal with the radiation terms left out, i.e.
	\begin{eqnarray}
		\frac{\partial (\rho\mathbf{u})}{\partial t} & = &  \lambda\nabla E_\text{r},\\
		\frac{\partial E_\text{r}}{\partial t} & = & \left(1-\frac{\kappa_{\rm P}}{\chi_{\rm R}}\right)\lambda\mathbf{u}\cdot\nabla E_\text{r},
	\end{eqnarray}
	\noindent where we have included an extra term in the radiation energy density due its transformation into a lab frame quantity (see Eq.~[\ref{eq:frame}]).
	These terms are treated as source terms on the right-hand side and are evolved through a simple explicit integration scheme.\\
	
	Now, the only terms left are the ones related to radiation diffusion, absorption, and emission. 
	These processes are governed by the equations
	\begin{eqnarray}
		\frac{\partial e}{\partial t} 	&=& -c\kappa_{\rm P}\left(a_{\rm r}T^4 - E_\text{r} \right), \label{eq:interal}\\
		\frac{\partial E_\text{r}}{\partial t}			&=& c\kappa_{\rm P}\left(a_{\rm r}T^4 - E_\text{r} \right) + \nabla\cdot\left(\frac{c\lambda}{\chi_{\rm R}}\nabla E_\text{r}\right), \label{eq:diffusion}
	\end{eqnarray}
	\noindent where $e$ is the internal energy density.
	These equations show that although the total energy remains constant the absorption and emission processes cause the exchange between radiation and thermal energy. 
	The second term on the right-hand side of the second equation is responsible for radiation diffusion. 
	To solve this system of equations, we follow an implicit scheme in order not to constrain the timestep too much as radiation processes are dictated by the speed of light and by the size of the grid cells. 
	The discretisation and solution of the system are developed in the next section.

\section{Implicit radiation solver}
\label{sec:app2}

	We follow the method developed by \cite{C11} for implementing FLD in RAMSES \citep{T02} as well as used in other codes such as PLUTO \citep{M07} by \cite{K13} and AZEuS \citep{R12} by \cite{R15}. 
	We start discretising Equations~(\ref{eq:interal}) and~(\ref{eq:diffusion}), applying $e=\rho C_{\rm V}T$, where $C_{\rm V}$ is the specific heat capacity at constant volume. 
	We also assume $\rho$ to be constant within the implicit step. 
	\begin{eqnarray}
		\rho C_{\rm V}\frac{T^{n+1}-T^n}{\Delta t} 	&=& -c\kappa_{\rm P}^n\left[a_{\rm r}(T^{n+1})^4 - E_\text{r}^{n+1} \right],\\
		\frac{E_\text{r}^{n+1}-E_\text{r}^n}{\Delta t}			&=& c\kappa_{\rm P}^n\left[a_{\rm r}(T^{n+1})^4 - E_\text{r}^{n+1} \right]+ \nabla\cdot\left(\frac{c\lambda^n}{\chi_{\rm R}^n}\nabla E_\text{r}^{n+1}\right).
	\end{eqnarray}
	\noindent We linearised the system assuming temperature changes are small within each timestep, i.e.
	\begin{eqnarray}
		(T^{n+1})^4=(T^n)^4\left(1+\frac{T^{n+1}-T^n}{T^n}\right)^4\approx4(T^n)^3T^{n+1}-3(T^n)^4.
	\end{eqnarray}
	Applying this approximation to our system of equations, it is possible to find the following equation for the temperature
	\begin{eqnarray}
		T^{n+1} = \frac{3a_{\rm r}\kappa^n_{\rm P}c\Delta t(T^n)^4+\rho C_{\rm V}T^n+\kappa_{\rm P}^nc\Delta t E^{n+1}_\text{r}}{\rho C_{\rm V}+4a_{\rm r}\kappa_{\rm P}^n c\Delta t(T^n)^3}.
		\label{eq:Tplus}
	\end{eqnarray}
	The radiation energy density equation is given by
	\begin{eqnarray}
		E^{n+1}_\text{r}-E^n_\text{r}- c\Delta t\nabla\cdot\left(\frac{\lambda^n}{\chi_{\rm R}^n}\nabla E_\text{r}^{n+1}\right)&=&c\Delta t\kappa_{\rm P}^n\left[4a_{\rm r}(T^n)^3T^{n+1}-3a_{\rm r}(T^n)^4-E^{n+1}_\text{r}\right]. \label{eq:Eplus}
	\end{eqnarray}
	Combining these two expressions it is possible to build a single linear equation where the unknowns are the radiation energy density in the next timestep, i.e. at $n+1$, 
	\begin{eqnarray}
		F(E^{n+1}_\text{r}) &= &c\Delta t\kappa_{\rm P}^n\left(4a_{\rm r}(T^n)^3\left[\frac{3a_{\rm r}\kappa^n_{\rm P}c\Delta t(T^n)^4+\rho C_{\rm V}T^n}{\rho C_{\rm V}+4a_{\rm r}\kappa_{\rm P}^n c\Delta t(T^n)^3}\right]-3a_{\rm r}(T^n)^4\right)+E^n_\text{r}.
		\label{eq:Fplus}
	\end{eqnarray}
	Here, $F=F(E_\text{r}^{n+1})$ includes all the terms with dependence of $E_\text{r}^{n+1}$ from Equation~\ref{eq:Eplus}, i.e. the left-hand side plus an extra term $g^n E^{n+1}_\text{r}$ that appears from the substitution of Equation~\ref{eq:Tplus}, where $g^n$ is known and given by
	\begin{eqnarray}
	    g^n=\frac{\rho C_\text{V}\kappa_\text{P}^n c\Delta t}{\rho C_\text{V}+4a_\text{r}\kappa^n_\text{P}c\Delta t(T^n)^3}.
	\end{eqnarray}
	\noindent As all terms on the right-hand side of Equation~\ref{eq:Fplus} are known we collectively label them as $b^n$. 
	
	Now we proceed to discretise the spatial derivatives. 
	Let us consider an arbitrary cell of a three-dimensional grid with indices $(i,j,k)$. 
	We integrate across the volume of such a cell $V_{i,j,k}$.
	\begin{eqnarray}
		\iiint\displaylimits_{V_{i,j,k}}F(E^{n+1}_\text{r})dV & = & \iiint\displaylimits_{V_{i,j,k}}b^n dV,\\
		\iiint\displaylimits_{V_{i,j,k}}(1+g^n)E^{n+1}_\text{r}- c\Delta t\nabla\cdot\left(\frac{\lambda^n}{\chi_{\rm R}^n}\nabla E_\text{r}^{n+1}\right)dV & = & b^nV_{i,j,k},\\
		(1+g^n)E^{n+1}_{r_{i,j,k}}-\frac{c\Delta t}{V_{i,j,k}}\oiint\displaylimits_{A(V_{i,j,k})}\frac{\lambda^n}{\chi_{\rm R}^n}\nabla E_\text{r}^{n+1}\cdot d{\bf A} & = & b^n,\\
		(1+g^n)E^{n+1}_{r_{i,j,k}}-\frac{c\Delta t}{V_{i,j,k}}\sum_{l}^{\rm faces}\frac{\lambda^n_l}{\chi_{{\rm R}_l}^n}\nabla E_{r_l}^{n+1}\cdot {\bf A}_{l} & = & b^n, \label{eq:linsys}
	\end{eqnarray}
	\noindent where the index $l$ runs for each face of the cell $(i,j,k)$, therefore quantities with the subscript $l$ are evaluated at the interface between neighbouring cells. 
	In a uniform Cartesian grid the number of faces is six but in this case the situation is not as simple. 
	First, the code JET solves the hydrodynamics equations in spherical coordinates and, more importantly, its moving-mesh nature allows the motion of cells in the radial direction. 
	The latter feature causes that any given cell, despite having two radial neighbouring cells, to have an arbitrary number of angular neighbouring cells, in both the polar and azimuthal directions. 
	Thus the index $l$ runs for an arbitrary number of faces, and the face area must be carefully calculated in order to account for grid misalignment of neighbouring cells.
	
	Once we have evaluated Equation~(\ref{eq:linsys}) for every single cell in the domain, we construct a set of linear equations defined by an $N\times N$ matrix, where $N$ is the number of cells in the grid at a given iteration\footnote{In JET the number of radial cells can change during the simulation}. 
	In order to solve the system we make use of the Portable, Extensible Toolkit for Scientific Computation \citep[PETSc;][]{B20}. 
	Unlike other implementations in grid-based codes (e.g. PLUTO, AZeUS, RAMSES), the matrix associated with the linear system is (in general) non-symmetric. 
	Thus, we cannot relay on using simple solvers such as the Conjugate Gradient method (CG), the Generalised Minimal Residual (GMRES). 
	In fact they neither show good convergence nor performance. 
	Instead, our tests showed that the Stabilised Bi-Conjugate Gradient Stable linear solver \citep[BCGS;][]{V92} together with a Block Jacobi (BJACOBI) preconditioner converges to the same solution as using a direct solver such as LU factorisation.  
	The use of relative and absolute convergence tolerance of $10^{-8}$ and $10^{-50}$, respectively, shows very good results in all our test cases (see Appendix~\ref{sec:app3}). 
	Nevertheless, it is important to remark that the choice of the appropriate combination of iterative linear solver and preconditioner seems to be problem dependent. 
	For instance, in the case of more complicated cases like the wind-reprocessed transient problem, these choices do not even converge. 
	Instead, a combination of BCGS together with an algebraic multigrid method (AMG) converges in a couple of iterations. 
	Specifically, we noticed that the use of the preconditioner boomerAMG \citep{Y02} from the library \textit{hypre} \citep{F06}, although more expensive computationally, manages to preserve stability and aids convergence.
	
\section{Radiation hydrodynamics validation tests}
\label{sec:app3}

	\begin{figure*}
	   \centering
	    \includegraphics[width=0.45\textwidth]{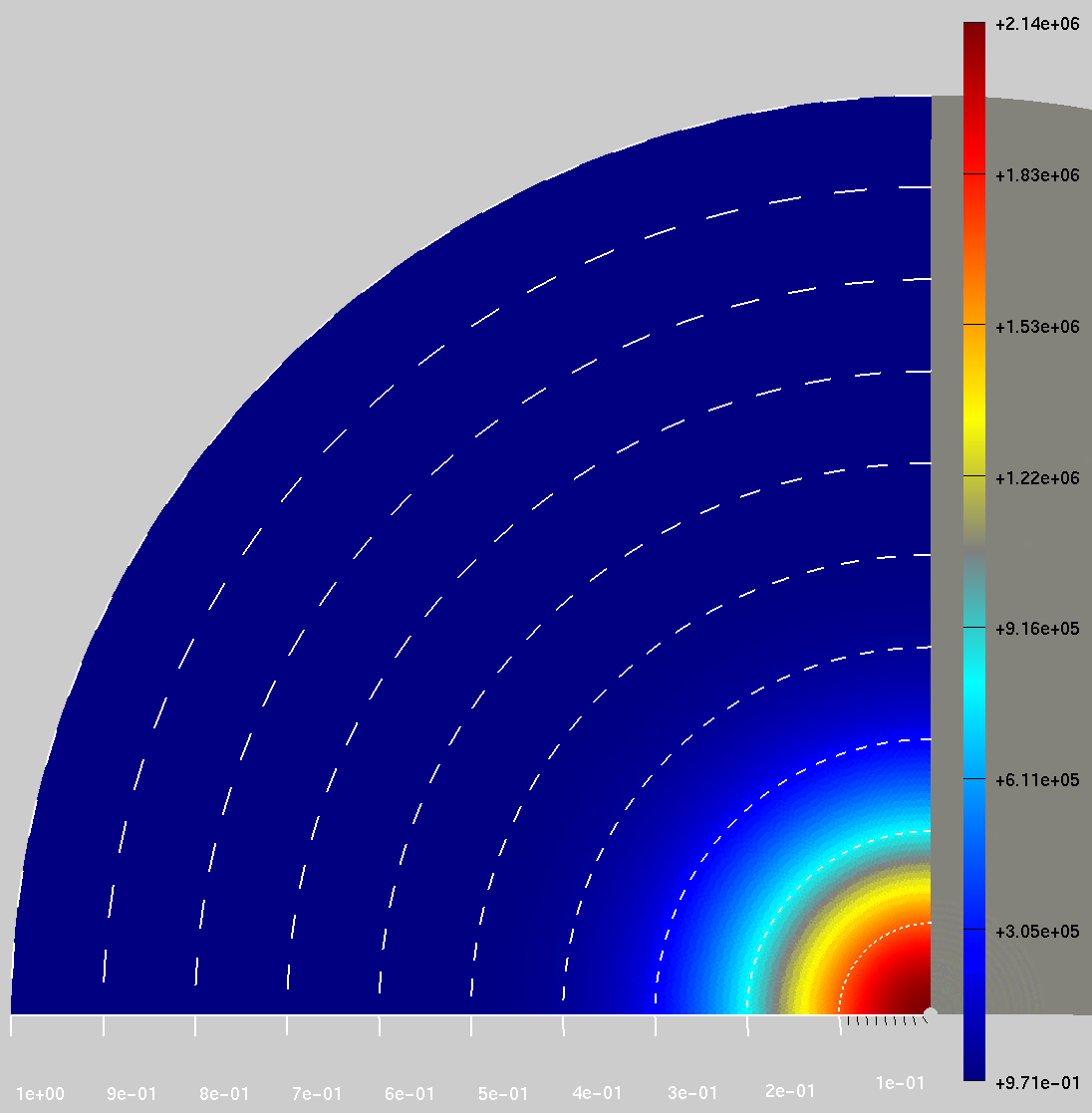} 
	    \hspace{0.06\textwidth}
	    \includegraphics[width=0.39\textwidth]{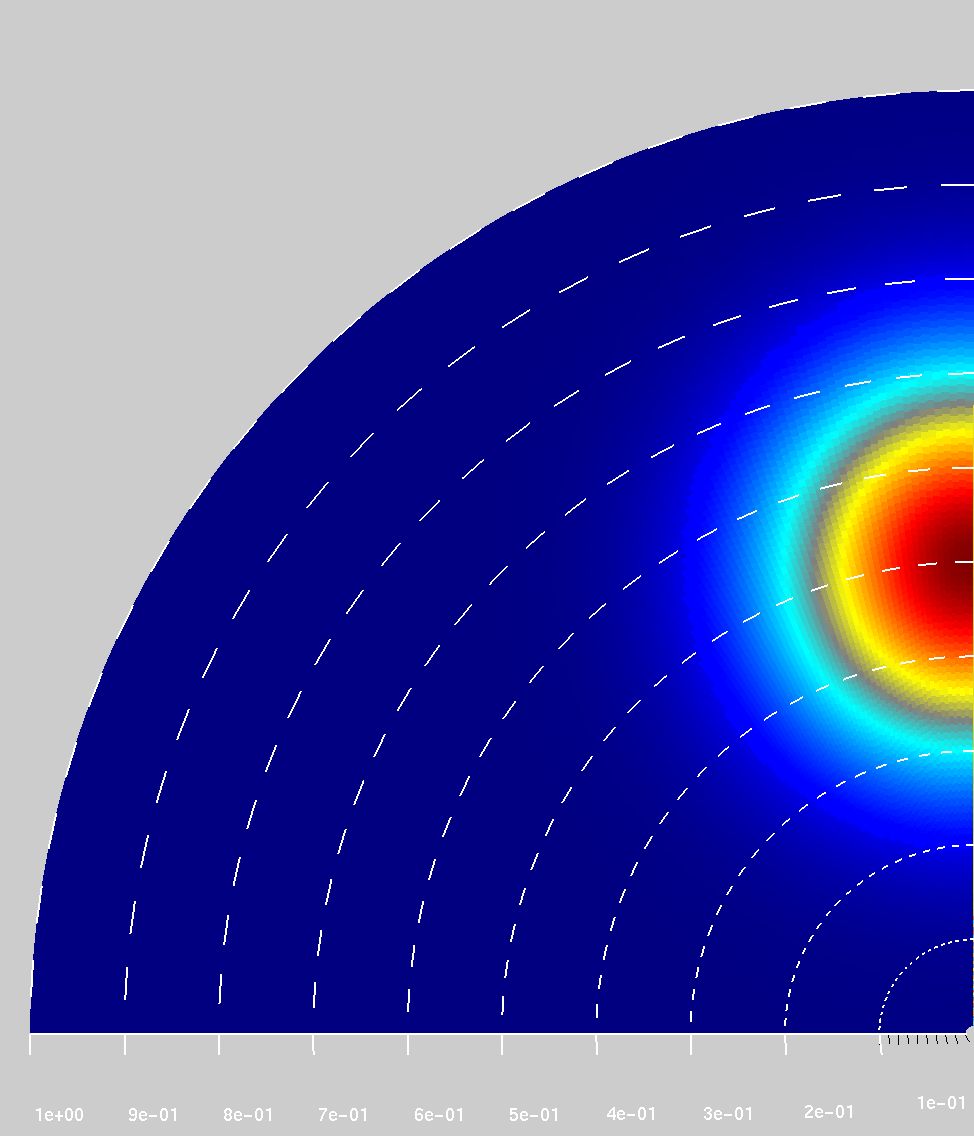} 
	   \includegraphics[width=0.495\textwidth]{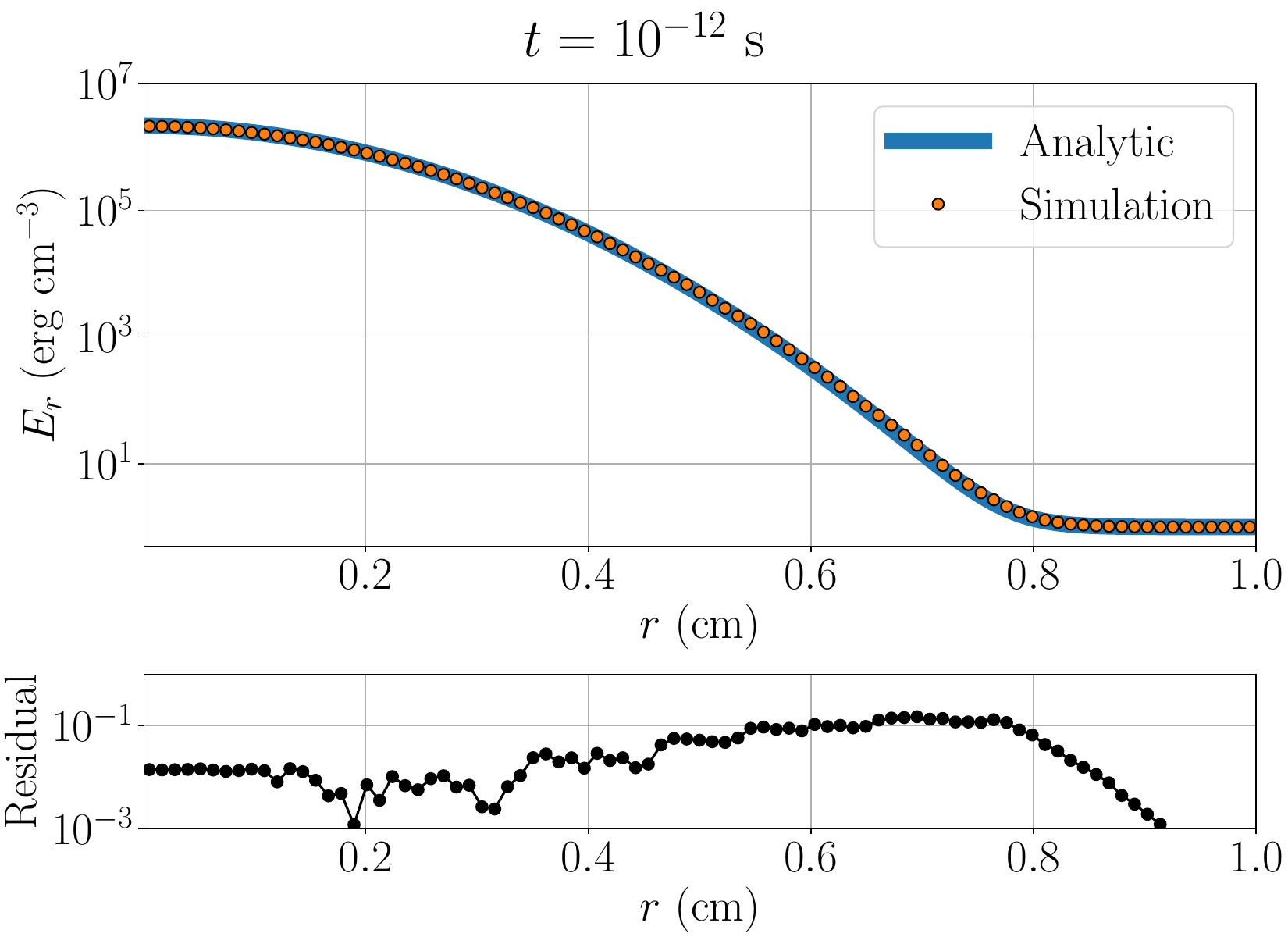} 
	   \includegraphics[width=0.495\textwidth]{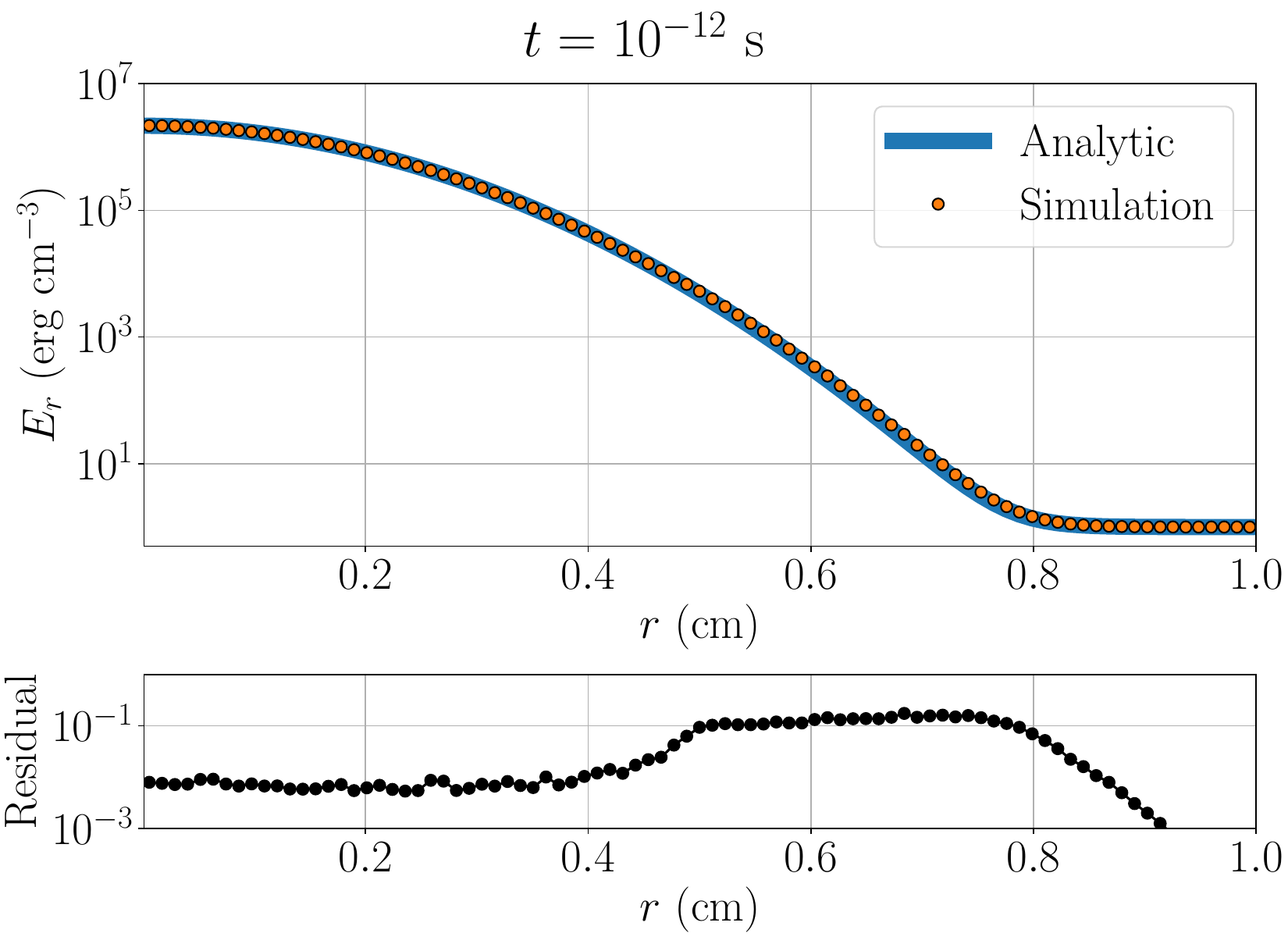} 
	   \caption{
	   Results of the radiation linear diffusion tests with the initial profile centred at the origin $r=0$ (left-hand side column), and at $r=0.5\rm\ cm$ (right-hand side column).
	   Upper panels show radiation density maps of the entire simulation domain at $t=10^{-12}\rm\ s$. 
	   Lower panels show radial profiles of the radiation energy density centred at $r=0$ and $r=0.5\rm\ cm$. 
	   The numerical solution (orange dots) is compared with the analytical solution (solid blue line). 
	   The residuals are shown with black dots connected with solid black lines. 
	   Differences are at most of $\sim10\%$, meaning the code models radiation diffusion correctly.
	   }
	   \label{fig:diffuse}
	\end{figure*}
	
	In order to validate the implementation of radiation into the hydrodynamics code JET, we performed simple tests problems with known and sometimes analytic solutions. 
	Inspired by previous works with similar implementations \citep[e.g.][]{C11,Z11, K13,R15,C19}, we present three tests in order to check different radiative process regimes. 

	\subsection{Linear diffusion}
	
		This test is based on the linear diffusion test by \cite{C11}. 
		Originally, this simulation recovers the analytical solution of a simple one-dimensional diffusion problem. 
		It starts with a delta function profile for the radiation energy density and evolves it.
		The hydrodynamic module of the code was disabled as this test aims to check the implementation of the radiation diffusion module alone. 
		In our case, we make use of a two-dimensional spherical grid but we did not start with a $\delta$ function profile. 
		Instead, the initial condition uses the analytical solution of the diffusion problem but short time after its initialisation as a $\delta$ function. 
		By doing so, we can avoid possible problems in the singularity of the spherical domain. 
		The profile of the radiation energy density is given by
		\begin{eqnarray}
		    E_\text{r}(r,t)=\frac{E_{\text{r},0}}{\sqrt{c\pi t}}\exp\left(-\frac{3r^2}{4ct}\right),
		\end{eqnarray}
		\noindent
		where $E_{\text{r},0}$ corresponds to the total radiation energy (i.e. the radiation energy density integrated over the domain). 
		We use a constant flux-limiter $\lambda = 1/3$, set $E_{\text{r},0}=10^5\rm\ erg\ cm^{-3}$, $\chi_\text{R}=1\rm\ cm^{-1}$, and $\kappa_{\rm P}=0$. 
		The domain in the radial and polar coordinates span $0\le r\le 1\rm\ cm$ and $0\le \theta\le 90\degr$ with $128\times128$ cells. 
		The density, pressure, and velocity were initialised to $\rho = 1\rm\ g\ cm^{-3}$, $p = 1\rm\ dyn\ cm^{-2}$, and ${\bf u} = {\bf 0}$, respectively. 
		In Figure~\ref{fig:diffuse}, specifically on the left-hand side column, we show the result of this simulation. 
		The top panel is a radiation energy density map of the entire domain at $t=10^{-12}\rm\ s$. 
		The lower panel shows a comparison between the analytical (solid blue line) and the numerical (orange dots) radiation energy density radial profiles at the same simulation time. 
		The residuals (connected black dots) are also included, showing that differences are kept $\lesssim10\%$. 
		This result implies that the code models radiation diffusion in the radial direction very well.
		
	\subsection{Off-center linear diffusion}
	
		For testing further the radiation diffusion module of the code we repeated the test of linear diffusion. 
		We use the same specifications and initial conditions.
		However, in this case the radiation energy density initial profile was not centred at the origin but at $r=0.5\rm\ cm$ and $\theta=0$ instead. 
		The result and comparison with the analytical solution are shown in the right column of Figure~\ref{fig:diffuse}. 
		Notice that the numerical solution matches the theoretical solution very well. 
		The residual are also of $\lesssim10\%$.
		This means that the radiation module handles diffusion correctly in the lateral direction too.
		
	\begin{figure}
	   \centering
	    \includegraphics[width=0.495\textwidth]{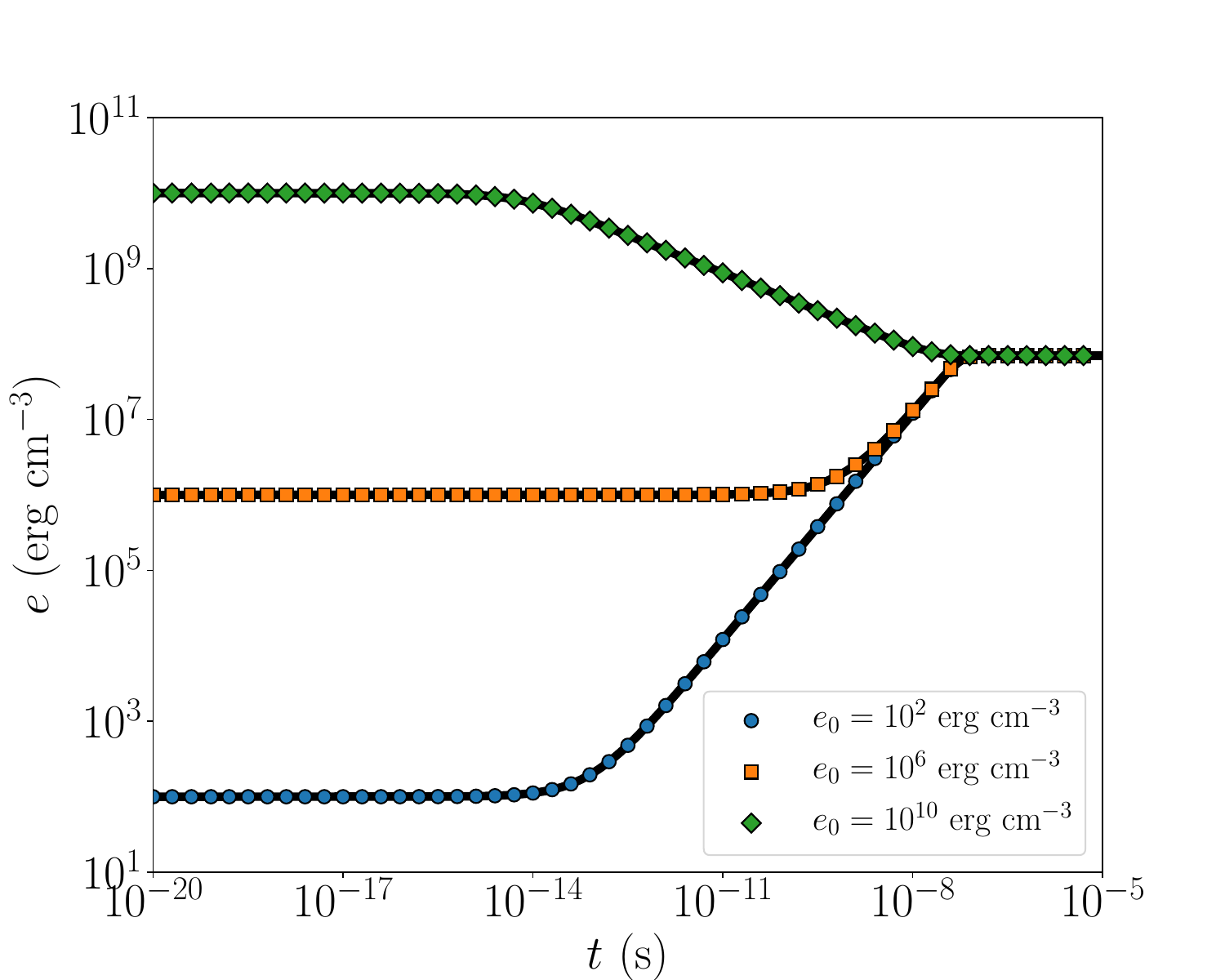} 
	   \caption{
	   Results of the matter-radiation coupling test. 
	   The initial internal energy density of the models was set to $e_0=10^{2}\,\rm erg\ cm^{-3}$ (blue dots), $10^{6}\,\rm erg\ cm^{-3}$ (orange squares), and $10^{10}\,\rm erg\ cm^{-3}$ (green diamonds). 
	   The analytical solution is also showed as a solid black line. 
	   The agreement between the analytical and numerical solution is excellent.
	   }
	   \label{fig:thermal}
	\end{figure}
	
	\begin{figure}
	   \centering
	    \includegraphics[width=0.495\textwidth]{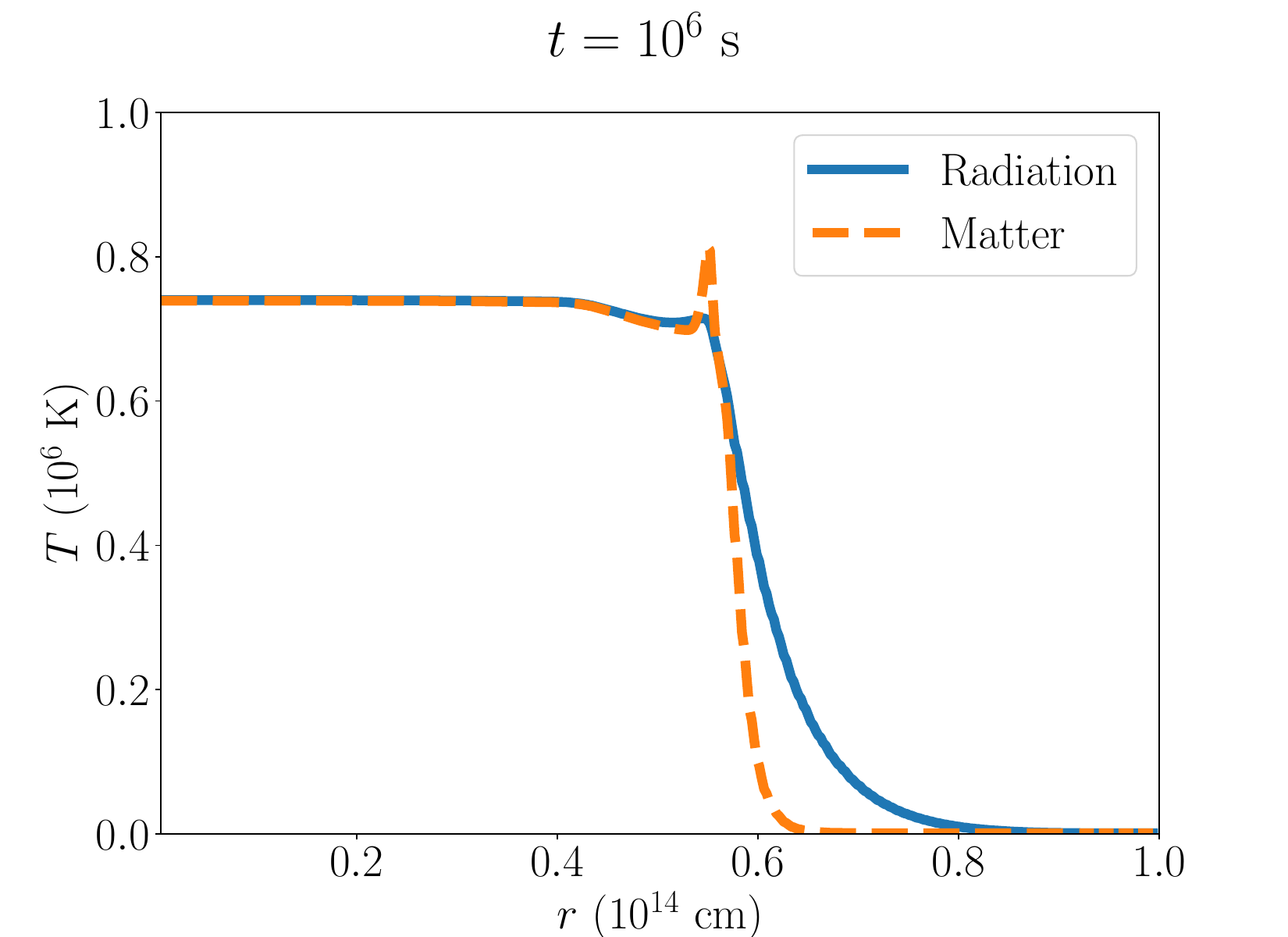}
	    \includegraphics[width=0.495\textwidth]{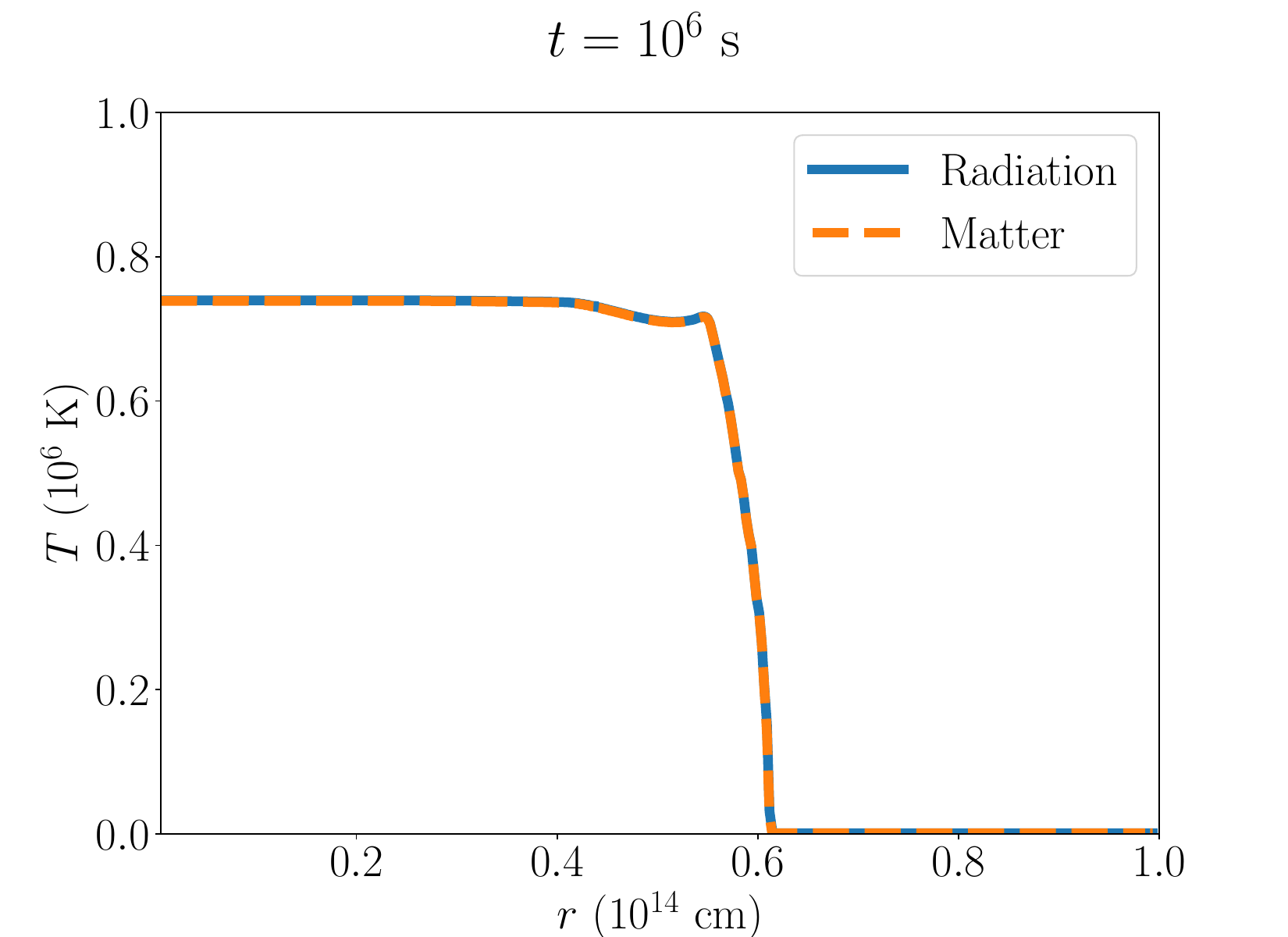} 
	   \caption{
	   Results of the radiative blast wave test for decoupled (upper panel) and coupled (lower panel) cases. 
	   Radiation (solid blue line) and matter (dashed orange line) temperature radial profiles at $t=10^6\rm\ s$. 
	   Notice that if radiation is not strongly coupled to matter it can diffuse forward, while gas temperature shows a peak due to the presence of a shock. 
	   }
	   \label{fig:decoupled}
	\end{figure}
			
	\subsection{Matter-radiation coupling}
	
		\cite{T01} proposed this test to check the coupling between matter and radiation. 
		The simulation considers a fluid at rest that it is not in thermal equilibrium. 
		Initially, most of the energy is in form of radiation, being constant across the domain. 
		Neglecting the importance of hydrodynamics it is possible to model the evolution of the internal energy density of the system as follows
		\begin{eqnarray}
			\frac{de}{dt} = c\kappa_{\rm P}E_\text{r}-c\kappa_{\rm P}a_\text{r}\left(\frac{\gamma-1}{\rho}\frac{\mu m_{\rm H}}{k_{\rm B}}\right)^4e^4.
		\end{eqnarray}
		\noindent
		Although this equation does not have an analytical solution, it is straightforward to solve it numerically. 
		The terms on the right-hand side represent the absorption and emission processes. 
		Both terms are equal in thermal equilibrium, which gives $\frac{de}{dt}=0$. 
		Finally, both the radiation and matter temperature should reach the same value, where the radiation temperature is defined by $T_\text{r}=\left(E_\text{r}/a_\text{r}\right)^{1/4}$.
		
		In this test, we only made use of the radiation module of the code by deactivating the hydrodynamic module of the code.
		The setup fills the entire domain with constant radiation energy density $E_\text{r}=10^{12}\rm\ erg\ cm^{-3}$, density $\rho=10^{-7}\rm\ g\ cm^{-3}$, Planck mean coefficient $\kappa_{\rm P} = 4\times10^8\rm\ cm^{-1}$, mean molecular weight $\mu=0.6$, and adiabatic index $\gamma=5/3$. 
		We consider a domain with $0 \le r \le 1$\,cm and $0\le \theta \le 90\degr$ with $32\times32$ cells. 
		We run three simulations, each of them with a different initial internal energy density: $e_0=10^{2}\rm\ erg\ cm^{-3}$, $10^{6}\rm\ erg\ cm^{-3}$, and $10^{10}\rm\ erg\ cm^{-3}$.
		The simulations start from $t=0$ and evolve up to $10^{-4}\,\rm\ s$, making sure the time sampling is sufficient for capturing the entire evolution. 
		As in \cite{K13}, we run three simulations with different values for the initial internal energy. 
		Figure~\ref{fig:thermal} shows the results of these tests along with the analytic solutions. 
		%The analytical solution is shown with a solid black line, while the numerical solutions of each model are shown in blue dots for $e_0=10^{2}\rm\ erg\ cm^{-3}$, orange squares for $e_0=10^{6}\rm\ erg\ cm^{-3}$, and green diamonds for $e_0=10^{10}\rm\ erg\ cm^{-3}$. 
		The agreement between the analytical (solid black lines) and numerical (blue circles, orange squares, and green diamonds) solutions is excellent. 
		Thus, the radiation module of the code can simulate the matter-radiation coupling very well. 
	
	\subsection{Radiative Blast Wave}
	
		This test simulates the evolution of a large amount of energy placed in a small region, which causes a spherical blast wave.  
		It is analogous to the purely hydrodynamic case of a Sedov-Taylor blast wave but it considers a high amount of radiation energy placed in the domain instead of thermal energy. 
		Despite not having an analytical solutions, this problem has been modelled by many codes as well as in different radiation-matter coupling regimes. 
		We follow the specifications by \cite{Z11} to set up the problem. 
		The domain spans $0 \le r \le 10^{14}\rm\ cm$ and $0 \le \theta \le 90\degr$. 
		The zones are linearly spaced.
		The medium is initialised with a constant density $\rho=5\times10^{-6}\rm\ g\ cm^{-3}$, at rest ${\bf u}={\bf 0}$, and with both radiation and matter temperature set to $T=10^3\rm\ K$. 
		However, in the central region of the domain, $r<2\times10^{12}\rm\ cm$, the radiation temperature is set to a higher value of $T_\text{r}=10^7\rm\ K$. 
		The adiabatic index is set to $\gamma=5/3$, and the mean molecular weight to $\mu=1$. 
		We tested two cases for different matter-radiation coupling regimes. 
		In the decoupled case, the Rosseland and Planck coefficients were set to $\kappa_{\rm P}=2\times10^{-16}\rm\ cm^{-1}$ and $\chi_{\rm R}=2\times10^{-10}\rm\ cm^{-1}$, respectively. 
		Meanwhile in the coupled case, both coefficients were set to $\kappa_{\rm P}=\chi_{\rm R}=2\times10^{-10}\rm\ cm^{-1}$.
		
		Figure~\ref{fig:decoupled} shows radial profiles for radiation (solid blue line) and matter (dashed orange line) temperatures at the end of the simulation at $t=10^6\rm\ s$ for the decoupled (upper panel) and coupled (lower panel) cases. 
		In the weak coupling case, we observe the presence of a sharp peak in the matter temperature as expected due to the presence of the shock. 
		The radiation can diffuse forward as it is not strongly coupled to the matter. 
		In the strong coupling case, both matter and radiation temperatures are identical. 
		Radiation is coupled to matter so it cannot diffuse ahead unlike in the previous case. 
		Both tests are in agreement with analogous tests by \cite{Z11} and \cite{C19}.

%%%%%%%%%%%%%%%%%%%%%%%%%%%%%%%%%%%%%%%%%%%%%%%%%%

% Don't change these lines
\bsp	% typesetting comment
\label{lastpage}
\end{document}